\documentclass[10pt,preprint2]{emulateapj}
\usepackage{color}
\usepackage{apjfonts}

\newcommand\kep{{\it Kepler}}
\newcommand{\blank}{\ensuremath{}}
\newcommand\muhz{\ensuremath{\mu\mathrm{Hz}}}
\newcommand\chisq{\ensuremath{\chi^2}}
\newcommand\ea{et al.}
\newcommand{\ph}{\ensuremath{\phantom{0}}}

\newcommand{\febyh}{\ensuremath{\mathrm{[Fe/H]}}}
\newcommand{\teff}{\ensuremath{T_\mathrm{eff}}}
\newcommand{\logg}{\ensuremath{\log g}}
\newcommand{\dzz}{\ensuremath{\Delta_0}}
\newcommand{\secdif}{\ensuremath{\Delta_2 \nu}}
\newcommand{\numax}{\ensuremath{\nu_\mathrm{max}}}
\newcommand{\BCZ}{\ensuremath{\mathrm{BCZ}}}
\newcommand{\HIZ}{\ensuremath{\mathrm{HeIIZ}}}
\newcommand{\tzero}{\ensuremath{T_0}}
\newcommand{\tbcz}{\ensuremath{T_\mathrm{BCZ}}}
\newcommand{\thiz}{\ensuremath{T_\mathrm{HeIIZ}}}
\newcommand{\taubcz}{\ensuremath{\tau_\mathrm{BCZ}}}
\newcommand{\tauhiz}{\ensuremath{\tau_\mathrm{HeIIZ}}}
\newcommand{\phibcz}{\ensuremath{\phi_\mathrm{BCZ}}}
\newcommand{\phihiz}{\ensuremath{\phi_\mathrm{HeIIZ}}}

\newcommand{\HeII}{He\,{\small II}}

\def\apj{ApJ}%
\def\apjl{ApJ}%
\def\apjs{ApJS}%
\def\apss{Ap\&SS}%
\def\aap{A\&A}%
\def\mnras{MNRAS}%

\def\myfigure#1#2#3#4{
	\begin{figure#4}
	\resizebox{\hsize}{!}{\includegraphics{#1}}
	\caption{#2 \label{#3}}
	\end{figure#4}}

\shorttitle{Acoustic glitches in solar-type stars}
\shortauthors{Mazumdar et al.}

\begin{document}


\title{Measurement of acoustic glitches in solar-type stars \\
       from oscillation frequencies observed by \kep}


\author{%
A.~Mazumdar\altaffilmark{1},
M.~J.~P.~F.~G.~Monteiro\altaffilmark{2,3},
J.~Ballot\altaffilmark{4,5},
H.~M.~Antia\altaffilmark{6},
S.~Basu\altaffilmark{7},
G.~Houdek\altaffilmark{8,9},
S.~Mathur\altaffilmark{10,15},
M.~S.~Cunha\altaffilmark{2},
V.~Silva Aguirre\altaffilmark{8,11},
R.~A.~Garc{\'{\i}}a\altaffilmark{12},
D.~Salabert\altaffilmark{13},
G.~A.~Verner\altaffilmark{14},
J.~Christensen-Dalsgaard\altaffilmark{8},
T.~S.~Metcalfe\altaffilmark{15,8},
D.~T.~Sanderfer\altaffilmark{16},
S.~E.~Seader\altaffilmark{17},
J.~C.~Smith\altaffilmark{17},
\and
W.~J.~Chaplin\altaffilmark{14}
}

\altaffiltext{1}{Homi Bhabha Centre for Science Education, TIFR, V.~N.~Purav Marg,
Mankhurd, Mumbai 400088, India}
\altaffiltext{2}{Centro de Astrof\'{\i}sica da Universidade do Porto, Rua das
Estrelas,4150-762 Porto, Portugal}
\altaffiltext{3}{Departamento de F\'{\i}sica e Astronomia, Faculdade de Ci\^encias
da Universidade do Porto, Rua do Campo Alegre, 4169-007 Porto,
Portugal}
\altaffiltext{4}{CNRS, Institut de Recherche en Astrophysique et
Plan\'etologie, 14 avenue Edouard Belin, 31400 Toulouse, France}
\altaffiltext{5}{Universit\'e de Toulouse, UPS-OMP, IRAP, 31400
Toulouse, France}
\altaffiltext{6}{Tata Institute of Fundamental Research, Homi Bhabha
Road, Mumbai 400005, India}
\altaffiltext{7}{Astronomy Department, Yale University, P.O. Box 208101,
New Haven, CT 065208101, USA}
\altaffiltext{8}{Stellar Astrophysics Centre, Department of Physics and
Astronomy, Aarhus University, Ny Munkegade 120, DK-8000 Aarhus C,
Denmark}
\altaffiltext{9}{Institute of Astronomy, University of Vienna, 1180,
Vienna, Austria}
\altaffiltext{10}{High Altitude Observatory, NCAR, P.O. Box 3000,
Boulder, CO 80307, USA}
\altaffiltext{11}{Max Planck Institut f\"ur Astrophysik,
Karl-Schwarzschild-Str. 1, 85748, Garching bei M\"{u}nchen, Germany}
\altaffiltext{12}{Laboratoire AIM, CEA/DSM, CNRS, Universit\'e Paris
Diderot, IRFU/SAp, Centre de Saclay, 91191 Gif-sur-Yvette Cedex, France}
\altaffiltext{13}{Laboratoire Lagrange, UMR7293, Universit\'e de Nice
Sophia-Antipolis, CNRS, Observatoire de la C\^ote d'Azur, 06304 Nice,
France}
\altaffiltext{14}{School of Physics and Astronomy, University of
Birmingham, Edgbaston, Birmingham, B15 2TT, UK}
\altaffiltext{15}{Space Science Institute, Boulder, CO 80301, USA}
\altaffiltext{16}{NASA Ames Research Center, Moffett Field, CA 94035, USA}
\altaffiltext{17}{SETI Institute/NASA Ames Research Center, Moffett
Field, CA 94035, USA}
          

\begin{abstract}
For the very best and brightest asteroseismic solar-type targets
observed by \kep, the frequency precision is sufficient to determine the
acoustic depths of the surface convective layer and the helium
ionization zone. Such sharp features inside the acoustic cavity of the
star, which we call acoustic glitches, 
create small oscillatory deviations from the uniform spacing of
frequencies in a sequence of oscillation modes with the same spherical
harmonic degree. We use these oscillatory signals to determine the
acoustic locations of such features in 19 solar-type stars observed by
the \kep\ mission. Four independent groups of researchers utilized the
oscillation frequencies themselves, the second differences of the
frequencies and the ratio of the small and large separation to locate
the base of the convection zone and the second helium ionization zone.
Despite the significantly different methods of analysis, 
good
agreement was found between the results of these four groups,
barring a few cases.
These results also agree reasonably well 
with
the locations of these layers in representative models of the stars.
These results firmly establish the presence of the oscillatory signals
in the asteroseismic data and the viability of several techniques to
determine the location of acoustic glitches inside stars.
\end{abstract}


\section{Introduction}
\label{sec:intro}

Acoustic glitches in a star are the regions where the sound speed
undergoes an abrupt variation due to a localized sharp change in the
stratification. The major acoustic glitches are the boundaries between
radiative and convective regions and the layers of ionization of
elements, especially hydrogen and helium. Such a glitch introduces an
oscillatory component, $\delta\nu$, in the eigenfrequencies of the star 
with respect to the frequencies themselves 
\citep{Gough88,Vorontsov88,Gough90}, proportional to
\begin{equation}
\delta\nu \propto \sin(4\pi\tau_\mathrm{g}\nu_{n,l} + \phi)\,,
\end{equation}
where
\begin{equation}
\tau_\mathrm{g}=\int^{R_\mathrm{s}}_{r_\mathrm{g}} {{\mathrm d}r\over
c}\,,
\end{equation}
is the acoustic depth of the glitch measured from the surface, $c$
the adiabatic sound speed, 
$r_\mathrm{g}$ the radial distance of the glitch,
$R_\mathrm{s}$ the seismic radius of the star (see discussion below),
$\nu_{n,l}$ the frequency of a mode with radial order $n$ and degree
$l$, and $\phi$ a phase factor. Each glitch will contribute to such a
signal in the frequencies with a ``periodicity'' of twice the acoustic
depth of the corresponding glitch. 

This oscillatory signature has been extensively studied for the
Sun in order to determine the extent of overshoot below the solar
convection zone
\citep{Vorontsov88,Gough90,Gough93,Monteiro94,Basu94,RV94,Basu97,JCD11}
and the seismic solar age \citep{HG11}. 

It has been proposed earlier that this may be used for distant stars also to find the position of the base of the convective envelope or the second helium ionization zone
\citep{Monteiro00,Mazumdar01,Gough02,RV03,Ballot04,Basu04,Houdek04,Mazumdar05,Piau05,HG06,HG07}. 
Indeed, \citet{Miglio10} have used the
modulation of the frequency separations to determine
the location of the second helium ionization zone in the red giant star,
HR7349, observed with the {\it CoRoT} satellite.
Recently, \citet{Mazumdar12} have used the oscillatory signal in the
second differences of the frequencies of the solar-type {\it CoRoT}
target star HD49933
to determine the acoustic depths of its second helium ionization
zone and the base of the convective envelope \citep[see
also][]{Mazumdar11,Rox11}. Earlier, \citet{Bedding10}
reported detection of an acoustic glitch in Procyon, although no
association to a specific layer was made. 

The scientific interest in studying the acoustic glitches goes beyond
the obvious goal of placing constraints on the positions of specific
layers in the stellar interior. The accurate determination of the
location and profile of the transition at the base of the convective envelope, for example, will help
us refine our understanding of the stellar dynamo in cool stars. On the
other hand, the amplitude of the oscillatory signal from the ionization
zones is directly related to the abundance of the corresponding element
in the star. For helium especially, an estimate of the helium abundance
in an ensemble of stars of different masses and ages will lead us to a
better understanding of the process of element enrichment in stars which
can be extrapolated back to the primordial helium content of the
universe, an important parameter in cosmology.

The acoustic glitches can be used to determine the acoustic depth of the
surface convection zone in a star, which in turn can be used to
constrain the stellar models. In particular, the position of the base of
the convection zone is very sensitive to opacity of stellar material,
which depends on the heavy element abundances. It is well known that the
recent determination of heavy element abundances \citep{Asplund09}
using 3D hydrodynamic atmospheric models for the Sun are not consistent
with helioseismic data \citep[][and references
therein]{Basu08,Gough13}. On the other hand, the older abundances of
\citet{GS98} estimated using 1D atmospheric models are
consistent with helioseismic data.  The cause of this discrepancy is not
understood and it would be interesting to test if asteroseismic data are
consistent with the revised abundances.  In particular, the heavy
element abundance, $Z$, for stars is estimated from observed \febyh\ 
ratio. Thus the estimated value of $Z$ depends on the heavy element
mixture (ratio of Fe to H abundance) for the Sun. Recently
\citet{vanSaders12}, noting the dependence on $Z$ of the acoustic depth 
of the convection zone, proposed that the measured acoustic depth can be 
used to constrain $Z$.

The effects of the acoustic glitches are subtle, and hence great care is
required in ascertaining that the results of the analyses reflect the
stellar properties at a significant level. Although analysis of
artificial data \citep[e.g.,][]{Basu04,Mazumdar05,mt05,HG07} are
useful in this regard, real data may well give rise to effects that are
not contained in such simulations. Here we test the reliability of the
inferences by comparing the results of several independent fits to the
glitch properties of the same frequency data, using rather different
techniques. These all provide measures of the acoustic locations of the
glitches which can be directly compared. Other properties of the fits,
however, depend more sensitively on the techniques and hence can only
properly be interpreted in the context of comparisons with model
results. We return to this point in Sect.~\ref{sec:summary}.

In the present analysis we studied 19 stars continuously observed by the
{\it Kepler mission} \citep{Borucki10,Koch10} during 9 months.
Launched on 2009 March 7, \kep\ is
monitoring 150\,000 stars in the constellations of Cygnus and Lyra
every half an
hour to look for Earth-like planets orbiting around solar-like stars.
Photometric time series of a subsample of 512 stars are studied at a
shorter cadence of 58.8\,s \citep{Gilliland10}. Every 3 months
the spacecraft rolls by $90^\circ$, to maintain the solar panels directed
towards the Sun. Therefore, the datasets are organized in quarters. The
time series of our sample of 19 stars were acquired during quarters 5 to
9, processed following the methods described by \citet{Garcia11}, and
their frequencies extracted as described by \citet{Appourchaux12}.

It must be noted that for the purpose of detection of the acoustic
glitches, it is sufficient to tag the frequencies only by their angular
degree, $l$. A visual inspection of the power spectrum yields a clear
identification of $l$ of the different modes. We do not need to know the
absolute overtone number, $n$, of the modes. We only need to ensure that
we do not have any ``missing'' orders (since the large separation, i.e.,
the average difference in frequency between successive overtones of the
same angular degree $l$, is known from the power spectrum itself, this
is a trivial step). Thus, when combining the frequencies to construct
the required diagnostics, if the angular degree is known, the correct
combination of frequencies can be made (in some cases that just means
taking differences of frequencies of modes of the same degree).  The
methods that we describe in the next section are, therefore, independent
of any input from stellar models which are usually used to identify the
radial order of observed modes. 

The choice of the sample of stars is only guided by the fact that these
were among the stars with the best signal to noise ratio in the \kep\
data. All the stars are classified as being either in
the main sequence or early sub-giant phase and thus have very few mixed
modes. Mixed modes hinder the detection of
the oscillatory signal of the acoustic glitches, and are best avoided
for the present purpose. There would be other similar
stars in the same region of the HR diagram, which are also expected to
exhibit similar signals of acoustic glitches; this is merely a random
sample.

In the next section we describe in detail the different techniques used
to determine the locations of the acoustic glitches. 
In Sect.~\ref{sec:results} we present the acoustic locations resulting 
from our fits to the observed frequencies, while
Sect.~\ref{sec:results_models}
compares them with analyses of associated stellar models. 
Sect.~\ref{sec:discussion} provides a discussion of results for 
individual stars.
We summarize the conclusions in
Sect.~\ref{sec:summary}.
Appendix~\ref{app:methods} presents details on the individual techniques, 
whereas
Appendix~\ref{app:model_details_YREC} provides details of the stellar models
used.

\section{The techniques}
\label{sec:technique}

We applied four different methods (labelled A to D in
Appendix~\ref{app:methods}) to determine
the acoustic locations of the base of the convective zone (\BCZ) and the
second helium ionization zone (\HIZ). The different methods were applied
to the same input data, namely the \kep\ frequencies of the stars, by
different subsets of the present authors independently, and the final
results are compared in Sec.~\ref{sec:results}. 

The first method (Method A) utilizes the oscillatory signals in the
frequencies themselves.  Method B fits a functional form to the second
differences of the frequencies 
with respect to the radial order
with acoustic depths of \BCZ\ and \HIZ\
among the free parameters. Method C fits the second differences of
the model frequency perturbations due to the glitches
to the observed second differences. Lastly, method
D utilizes the oscillatory signal present in the ratio of the small to
the large separation.
The large separation is the average difference in frequencies of same degree
and successive radial order, while the small separation is the difference
in frequency between a radial mode and the quadrupole mode of the
previous radial order. 
Methods~A, B, and C determine the acoustic depths (measured from the
surface) of the \BCZ\ and \HIZ\ while method~D determines the acoustic
radius (measured from the center) of the \BCZ.
The details of each method are described in Appendix~\ref{app:methods}.
We note that given that the methods employ different techniques on
different combinations of frequencies the systematic effects would be
quite different for each of them and it would be reasonable to expect
considerable scatter in the results to be found from them.

For the rest of the paper, we adopt the following definitions and 
notations for the acoustic radii and acoustic depth.
\tbcz\ and \thiz\ are the acoustic radii, defined as
\begin{equation}
\tbcz=\int_0^{r_\mathrm{\BCZ}} \frac{\mathrm{d}r}{c}\,;
\quad
\thiz=\int_0^{r_\mathrm{\HIZ}} \frac{\mathrm{d}r}{c}\,,
\label{eq:tbcz_thiz}
\end{equation}
where $r_\mathrm{\BCZ}$ and
$r_\mathrm{\HIZ}$ are the radial positions of the two glitches.
In particular, the radial position of the local minimum of the 
adiabatic index $\gamma_1= (\partial\ln
p/\partial\ln\rho)_s$ ($p$ and $\rho$ are pressure and density
respectively and $s$ is specific entropy) in the \ion{He}{2} region 
is taken to be $r_\mathrm{\HIZ}$.
The acoustic depths of \BCZ\ and \HIZ\ are \taubcz\ and \tauhiz, respectively, defined as
\[
\taubcz=\int^{R_\mathrm{s}}_{r_\mathrm{\BCZ}} \frac{\mathrm{d}r}{c}
\equiv \tzero - \tbcz\,;\nonumber
\]
\begin{equation}
\tauhiz=\int^{R_\mathrm{s}}_{r_\mathrm{\HIZ}} \frac{\mathrm{d}r}{c}
\equiv \tzero - \thiz\,.
\label{eq:taubcz_tauhiz}
\end{equation}
Here \tzero\ is the total acoustic radius of the star, which can
be calculated as 
\begin{equation}
\tzero=\int_0^{R_\mathrm{s}} \frac{\mathrm{d}r}{c}\,,
\label{eq:tzero}
\end{equation}
where $R_\mathrm{s}$ is the seismic radius of the star, which may be
considered as a fiducial radius
that defines the
outer phase of the acoustic modes, subject to the chosen boundary
conditions, relative to the phase in the propagating region below the
turning point. $R_{\mathrm s}$ can be determined from fitting an
approximated atmosphere, such as a polytropic atmosphere, to a more
realistic stellar atmosphere (such as that obtained 
from 3D numerical simulations or from observations).
In the case of the Sun this fiducial radius is above the temperature
minimum, i.e., rather far away from the photosphere \citep[about 225\,s
above the radius corresponding to the effective temperature in the Sun,
see][]{Monteiro94,HG07}. 
For the models used in this work (see Sec.~\ref{sec:results_models}), the outer boundary for the calculation of the total acoustic radius has been
assumed to be the surface of the star including the atmosphere.
The choice of the outer acoustic boundary may affect the value of the
acoustic depth as determined by any of these methods. 

The total acoustic radius can also be
estimated from the average large separation, \dzz, as
$\tzero \approx (2\dzz)^{-1}$.
The validity of this approximation has been studied by 
\citet{Hekker13}.
There are at least three different methods available in
the literature for determining the observed value of \dzz: (1) from
taking a mean or median value over some arbitrarily chosen frequency range for
modes of like degree $l$, (2) from determining the frequency of the
corresponding spectral peak in the Fourier spectrum of the power
spectrum of the observed oscillations, (3) from fitting the asymptotic
expression for solar-like oscillations \citep{Tassoul80,Gough86} to the
observed frequencies.
Here we calculated \dzz\ by fitting a linear
relation to the observed radial mode frequencies as a function of the
radial order, which is equivalent to the third method above.
In principle, an estimation of \tzero\ from \dzz\ measured by another
alternative technique would yield a slightly different value. 
However, given that the large separation is a fairly robust quantity, especially for stars with high signal to noise ratio \citep[see][]{Verner11}, we do not expect this to be a major source of uncertainty.

\begin{deluxetable}{lrrrr}
\tabletypesize{\small}
\tablecaption
{Basic spectroscopic and seismic data for 19 \kep\ stars
\label{tab:spec_seismo}
}
\tablewidth{0pt}

\tablehead{
\colhead{KIC ID} 
   & \colhead{\teff} 
   & \colhead{\logg} 
   & \colhead{\febyh} 
   & \colhead{{\dzz}}\\
& \colhead{(K)} &       &        & \colhead{(\muhz)}
}

\startdata

KIC008006161
& $ 5390$ & $ 4.47$ & $ 0.38$ & $ 149.1{\pm }  0.1$
\\

KIC008379927
& $ 5960$ & $ 4.39$ & $-0.30$ & $ 119.9{\pm }  0.1$
\\

KIC008760414
& $ 5787$ & $ 4.33$ & $-1.19$ & $ 117.0{\pm }  0.1$
\\

KIC006603624
& $ 5625$ & $ 4.31$ & $ 0.26$ & $ 109.9{\pm }  0.1$
\\

KIC010454113
& $ 6120$ & $ 4.32$ & $-0.07$ & $ 105.1{\pm }  0.2$
\\

KIC006106415
& $ 5990$ & $ 4.29$ & $-0.11$ & $ 103.7{\pm }  0.1$
\\

KIC010963065
& $ 6060$ & $ 4.28$ & $-0.21$ & $ 102.4{\pm }  0.1$
\\

KIC006116048
& $ 5935$ & $ 4.27$ & $-0.26$ & $ 100.3{\pm }  0.2$
\\

KIC004914923
& $ 5905$ & $ 4.19$ & $ 0.14$ & $  88.3{\pm }  0.1$
\\

KIC012009504
& $ 6065$ & $ 4.21$ & $-0.09$ & $  87.7{\pm }  0.1$
\\

KIC012258514
& $ 5990$ & $ 4.12$ & $ 0.02$ & $  74.5{\pm }  0.1$
\\

KIC006933899
& $ 5860$ & $ 4.08$ & $ 0.01$ & $  71.8{\pm }  0.1$
\\

KIC011244118
& $ 5745$ & $ 4.07$ & $ 0.34$ & $  71.3{\pm }  0.2$
\\

KIC008228742
& $ 6042$ & $ 4.02$ & $-0.15$ & $  61.6{\pm }  0.1$
\\

KIC003632418
& $ 6190$ & $ 4.01$ & $-0.19$ & $  60.5{\pm }  0.1$
\\

KIC010018963
& $ 6020$ & $ 3.95$ & $-0.47$ & $  55.2{\pm }  0.2$
\\

KIC007976303
& $ 6053$ & $ 3.90$ & $-0.52$ & $  50.9{\pm }  0.2$
\\

KIC011026764
& $ 5682$ & $ 3.89$ & $-0.26$ & $  50.2{\pm }  0.1$
\\

KIC011395018
& $ 5424$ & $ 3.84$ & $-0.39$ & $  47.4{\pm }  0.1$
\\

\enddata

\end{deluxetable}

\myfigure{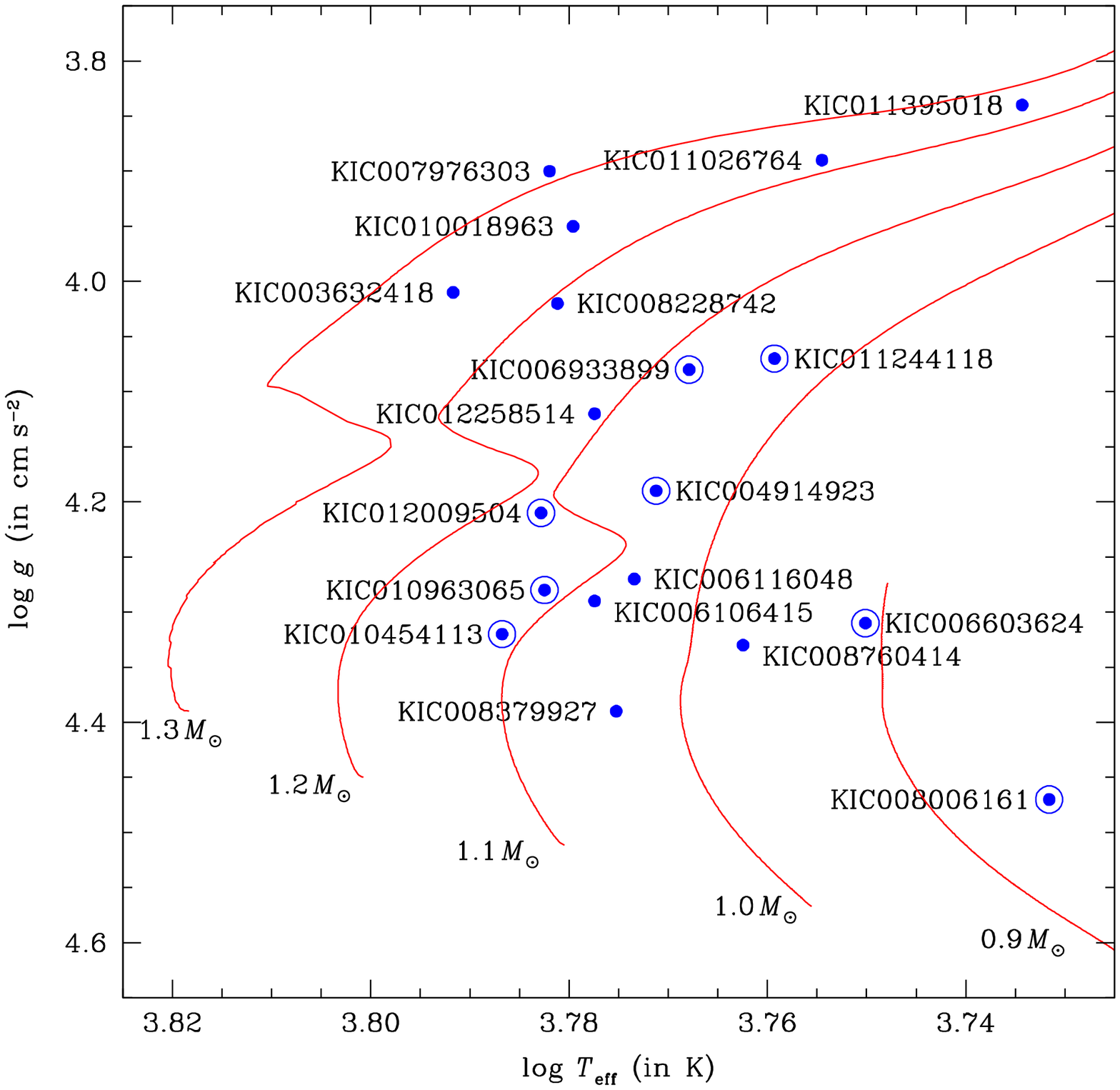}
{Hertzsprung-Russell diagram for 19 \kep\ stars. The eight stars
discussed in detail are circled. The {\it red} lines are evolutionary
tracks with solar chemical composition and indicated mass computed with
the CESAM2k code \citep{ml08}.}
{fig:hrd}
{} 

\section{Results}
\label{sec:results}

\myfigure{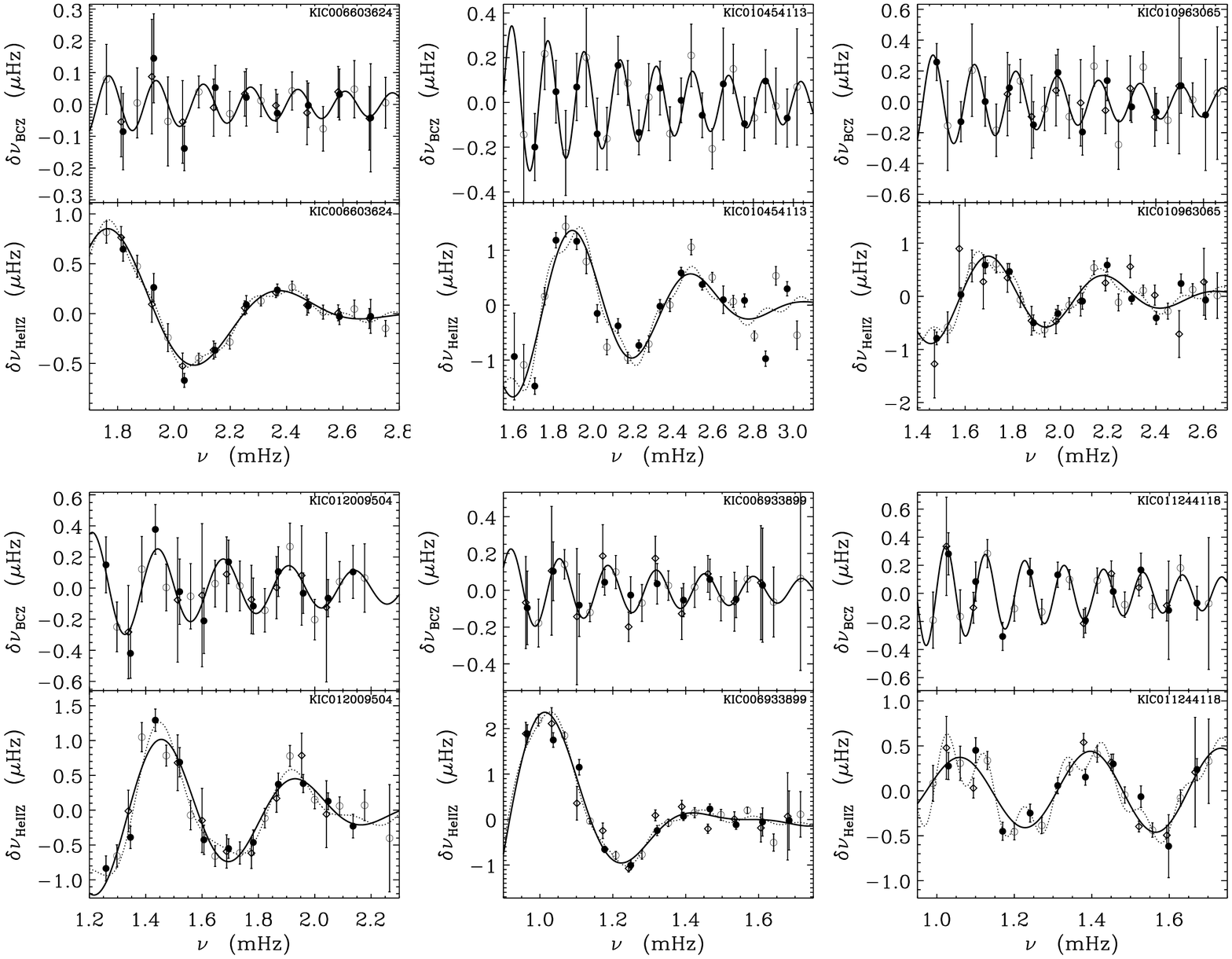}
{Illustration of method A: Fits of Eqs.~(\ref{eq:sig_mjm_bcz}) and
(\ref{eq:sig_mjm_hiz}) to the residuals of frequencies of six \kep\ stars
(KIC numbers given in each panel) after removing a smooth component iteratively. The points correspond to $l=0$ (black circles), $l=1$ (open circles) and $l=2$ (diamonds).
The {\it solid line} is the best fit to the \BCZ\ component (Eq.~(\ref{eq:sig_mjm_bcz}), {\it upper panel} for each star) and the \HIZ\ component (Eq.~(\ref{eq:sig_mjm_hiz}), {\it lower panel} for each star). 
In the {\it lower panels} the dotted line shows the \BCZ\ component superimposed on the \HIZ\ component, as obtained separately and shown in the upper panel.}
{fig:res_MJM}
{*}

\myfigure{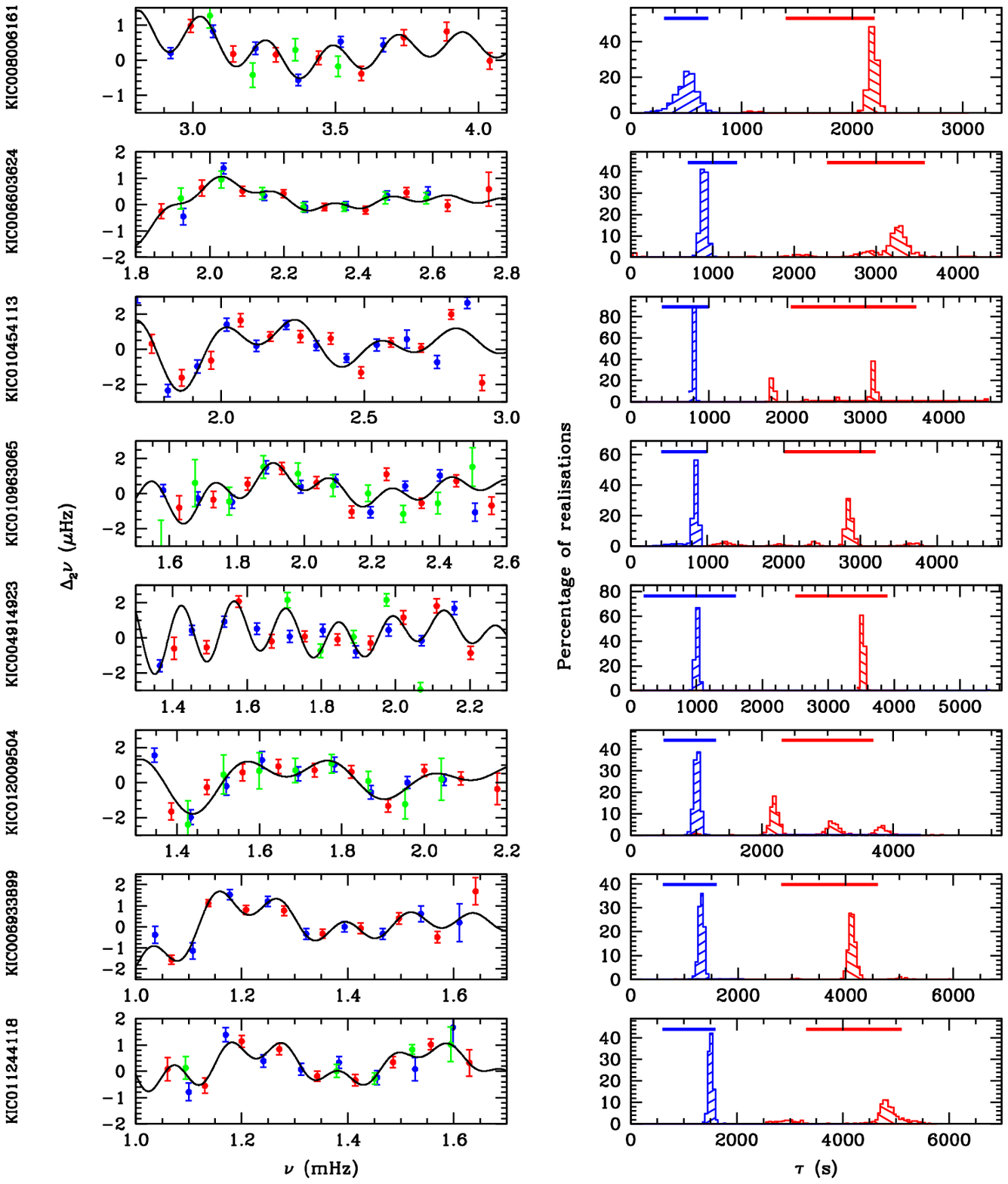}
{
Illustration of method~B: Fits of Eq.~(\ref{eq:hou08}) to the second 
differences of the mean frequencies for eight \kep\ stars (KIC numbers 
on the {\it extreme left}) and the histograms for the fitted values of
\tauhiz\ and \taubcz\ for different realizations of the data.
The second differences of the frequencies of $l=0$ ({\it blue}), 
$l=1$ ({\it red}) and $l=2$ ({\it green}) modes of the stars
and their fit to Eq.~(\ref{eq:hou08}) ({\it black curve}) are shown in
the {\it left} panels.  The
corresponding histograms of the fitted values of \taubcz\ (in {\it red})
and \tauhiz\ (in {\it blue}) for different realizations are shown in the
{\it right} panels.  The {\em solid} bands at the
top of the {\em right panels} indicate the range of initial guesses for
the two parameters in each fit.
}
{fig:res_ABM}
{*}

We applied the techniques described in Sect.~\ref{sec:technique} and
Appendix~\ref{app:methods} to 19
stars observed by the \kep\ mission. The frequencies of these stars were
given by \citet{Appourchaux12}. While methods A, B and D were applied
to all 19 stars, method C was applied to 
10 stars. The basic
spectroscopic parameters and the average large separation for the stars
are given in Table~\ref{tab:spec_seismo}. The effective temperatures
were adopted from \citet{Bruntt12} while the $\log g$ values were
determined from a seismic pipeline \citep{Basu10}. 
The positions of the 19 stars on the Hertzsprung-Russell diagram are shown in Fig.~\ref{fig:hrd}.
\myfigure{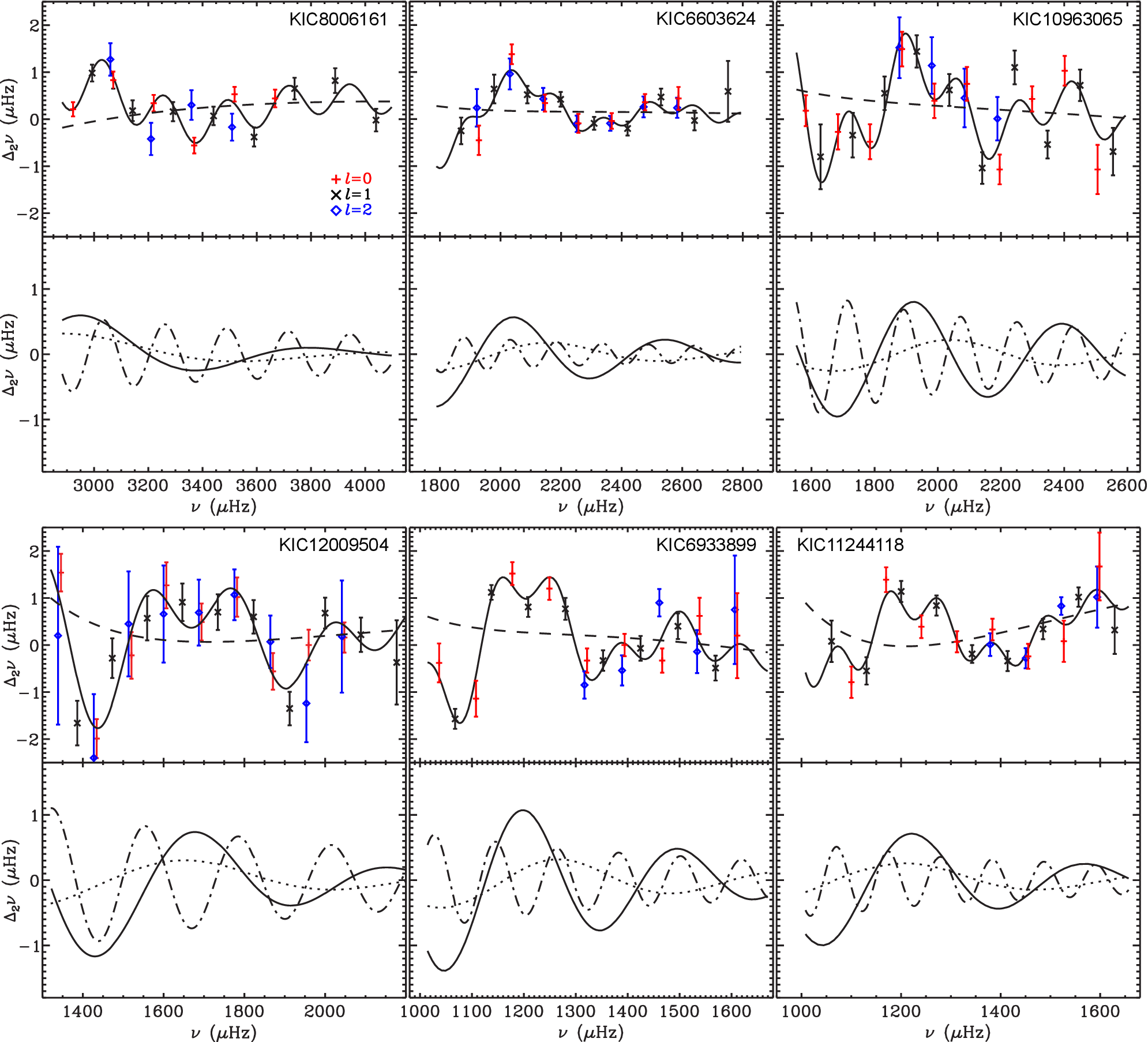}
{Illustration of method~C: Analyses results for 
six \kep\ stars (KIC numbers given in each panel).
The symbols in the {\it upper panels} denote second differences
\secdif\ for low-degree modes. The solid curves are fits to 
\secdif\ based on the analysis by 
\citet{HG07,HG11}. The 
dashed curves are the smooth contributions, including a third-order 
polynomial in $\nu^{-1}_i$ to represent the upper-glitch contribution
from near-surface effects. The {\it lower panels} display the 
remaining individual contributions from the acoustic glitches to 
\secdif: the {\em dotted} and {\em solid} curves are the contributions 
from the first and second stages of helium ionization, and the 
{\em dot-dashed} curve is the contribution from the acoustic glitch at the 
base of the convective envelope.}
{fig:res_GH}
{*}
\myfigure{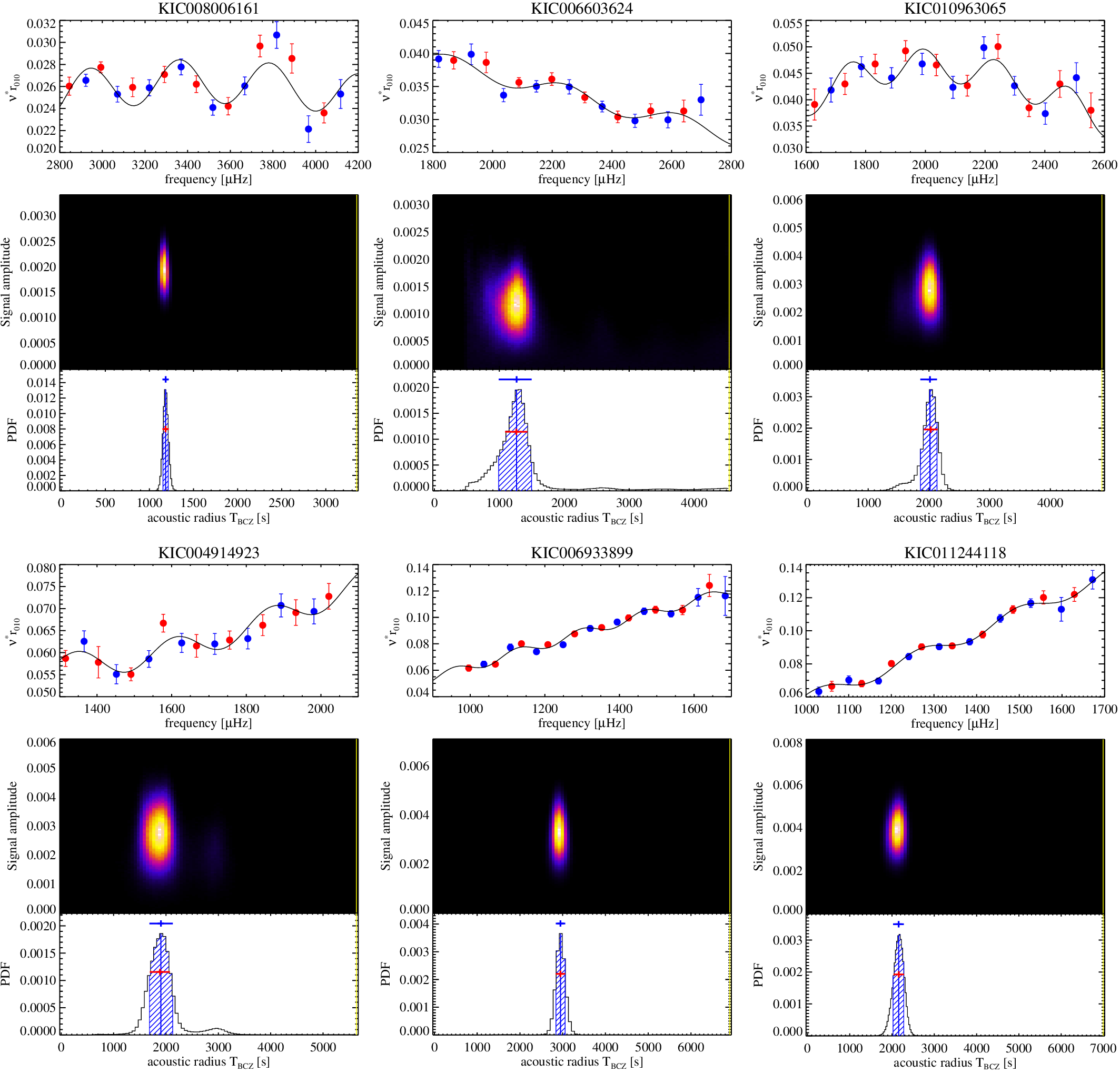}
{Illustration of method~D: Determination of \tbcz\ using frequency ratios 
for six \kep\ targets. Each of the six panels corresponds to a target and is
divided into three sub-panels. {\em Top} sub-panels show the seismic variables
$\nu^*r_{010}$. {\em Blue (red)} dots with error bars are the
observed values for $\nu^*r_{01}$ ($\nu^*r_{10}$). {\em Solid} lines show
models (Eq.~(\ref{eq:modd01})) with the parameters corresponding to the
highest posterior probabilities found by MCMC. {\em Middle} sub-panels are 2-D
probability functions in the plane ($T$,$A$). {\em White (black)} color
corresponds to high (low) probability. {\em Bottom} sub-panels show marginal
probability distributions for the parameter $T$. Vertical {\em blue} lines
indicate the medians of the distributions and hatched areas show the
68\%-level confidence intervals. {\em Blue} error bars, plotted above the
peaks, are uncertainties deduced from these intervals, whereas {\em red} uncertainties,
plotted across the peaks, are obtained by fitting the peaks with
Gaussian profiles.}
{fig:res_JB}
{*}
We show the results from each method graphically in
Figs.~\ref{fig:res_MJM}, \ref{fig:res_ABM}, \ref{fig:res_GH} and
\ref{fig:res_JB} for a selected set of eight stars: 
KIC008006161, KIC006603624, KIC010454113, KIC010963065,
KIC004914923, KIC012009504, KIC006933899, and KIC011244118.  
However, for methods A and D, not all the cases led to significant
results for acoustic depths or radii of \BCZ\ and \HIZ.

\begin{deluxetable*}{lllllllllll}
\tabletypesize{\small}

\tablecaption{Comparison of acoustic depths of the base of the convective
envelope (\taubcz) and the second helium ionization zone (\tauhiz) by
four independent methods.
\label{tab:res_comp}}

\tablewidth{0pt}

\tablehead{
\colhead{KIC ID} 
  & \multicolumn{1}{c}{\tzero~(s)} 
  & \colhead{\quad\quad}
  & \multicolumn{4}{c}{\taubcz~(s)} 
  & \colhead{\quad\quad}
  & \multicolumn{3}{c}{\tauhiz~(s)} \\
\colhead{}  
  & \colhead{} 
  & \colhead{}
  & \colhead{Method A} 
  & \colhead{Method B} 
  & \colhead{Method C} 
  & \colhead{Method D} 
  & \colhead{}
  & \colhead{Method A} 
  & \colhead{Method B} 
  & \colhead{Method C}
}

\startdata

KIC008006161*
& $\ph  3353 \pm    2$ &
& $  2284 ^{+  47} _{-  47}$
& $  2186 ^{+  35} _{-  39}$
& $  2199 ^{+  40} _{-  38}$
& $  2169 ^{+  30} _{-  30}$ &
& $\cdots$
& $\ph508 ^{+  80} _{- 101}$
& $\ph615 ^{+ 123} _{- 144}$
\\

KIC008379927
& $\ph  4170 \pm    3$ &
& $  1796 ^{+ 114} _{- 114}$
& $  1840 ^{+  29} _{-  26}$
& $  1858 ^{+  40} _{-  38}$
& $ (3172 ^{+  61} _{-  61})$ &
& $\cdots$
& $\ph714 ^{+  25} _{-  27}$
& $\ph827 ^{+  48} _{-  57}$
\\

KIC008760414
& $\ph  4273 \pm    3$ &
& $  2750 ^{+ 132} _{- 132}$
& $  2481 ^{+ 316} _{- 100}$
& $\blank$
& $  2430 ^{+ 141} _{- 141}$ &
& $\cdots$
& $\ph973 ^{+  82} _{- 185}$
& $\blank$
\\

KIC006603624*
& $\ph  4549 \pm    4$ &
& $  2932 ^{+ 183} _{- 183}$
& $  3234 ^{+ 133} _{- 438}$
& $  3285 ^{+ 130} _{- 196}$
& $  3287 ^{+ 171} _{- 171}$ &
& $\ph829 ^{+  89} _{-  89}$
& $\ph897 ^{+  36} _{-  37}$
& $  1031 ^{+  61} _{-  96}$
\\

KIC010454113*
& $\ph  4757 \pm    9$ &
& $  2801 ^{+ 125} _{- 125}$
& $  3083 ^{+  33} _{- 154}$
& $  3126 ^{+  75} _{- 122}$
& $\cdots$ &
& $\ph823 ^{+  18} _{-  18}$
& $\ph804 ^{+  13} _{-  15}$
& $\ph840 ^{+   7} _{-   8}$
\\

KIC006106415
& $\ph  4821 \pm    4$ &
& $  2903 ^{+ 242} _{- 242}$
& $  2875 ^{+ 393} _{- 247}$
& $\blank$
& $  2811 ^{+ 118} _{- 118}$ &
& $\cdots$
& $\ph908 ^{+  74} _{-  75}$
& $\blank$
\\

KIC010963065*
& $\ph  4882 \pm    4$ &
& $  2832 ^{+ 155} _{- 155}$
& $  2854 ^{+  47} _{-  42}$
& $  2803 ^{+ 120} _{- 166}$
& $  2852 ^{+ 112} _{- 112}$ &
& $  1020 ^{+  62} _{-  62}$
& $\ph851 ^{+  24} _{-  28}$
& $  1101 ^{+  27} _{-  29}$
\\

KIC006116048
& $\ph  4985 \pm    9$ &
& $  3153 ^{+ 220} _{- 220}$
& $  2942 ^{+ 178} _{- 273}$
& $  3003 ^{+ 141} _{- 347}$
& $  3134 ^{+ 186} _{- 186}$ &
& $  1015 ^{+ 108} _{- 108}$
& $  1048 ^{+  34} _{-  35}$
& $  1094 ^{+  11} _{-  92}$
\\

KIC004914923*
& $\ph  5662 \pm    6$ &
& $  3744 ^{+  91} _{-  91}$
& $  3525 ^{+  23} _{-  21}$
& $  3526 ^{+  33} _{-  29}$
& $  3777 ^{+ 191} _{- 191}$ &
& $\cdots$
& $  1006 ^{+  25} _{-  25}$
& $  1151 ^{+   7} _{-  56}$
\\

KIC012009504*
& $\ph  5701 \pm    6$ &
& $  2150 ^{+  74} _{-  74}$
& $  2174 ^{+  90} _{- 288}$
& $  2192 ^{+  84} _{- 103}$
& $\cdots$ &
& $  1039 ^{+  47} _{-  47}$
& $  1004 ^{+  40} _{-  47}$
& $  1089 ^{+  12} _{-  66}$
\\

KIC012258514
& $\ph  6711 \pm    9$ &
& $  2645 ^{+ 151} _{- 151}$
& $  2819 ^{+ 173} _{- 338}$
& $\blank$
& $\cdots$ &
& $  1277 ^{+  81} _{-  81}$
& $  1250 ^{+  67} _{-  78}$
& $\blank$
\\

KIC006933899*
& $\ph  6963 \pm    9$ &
& $  3773 ^{+ 201} _{- 201}$
& $  4115 ^{+  74} _{-  65}$
& $  4228 ^{+ 675} _{- 143}$
& $  4014 ^{+ 107} _{- 107}$ &
& $  1114 ^{+ 124} _{- 124}$
& $  1315 ^{+  54} _{-  45}$
& $  1721 ^{+  77} _{-  64}$
\\

KIC011244118*
& $\ph  7012 \pm   19$ &
& $  4829 ^{+ 251} _{- 251}$
& $  4844 ^{+ 246} _{- 133}$
& $  4798 ^{+ 165} _{- 576}$
& $  4851 ^{+ 125} _{- 125}$ &
& $  1490 ^{+  40} _{-  40}$
& $  1506 ^{+  41} _{-  33}$
& $  1501 ^{+  88} _{- 132}$
\\

KIC008228742
& $\ph  8116 \pm   13$ &
& $  4409 ^{+ 290} _{- 290}$
& $  4468 ^{+  89} _{- 166}$
& $\blank$ 
& $  4565 ^{+ 187} _{- 187}$ & 
& $\cdots$
& $  1509 ^{+  38} _{-  46}$
& $\blank$
\\

KIC003632418
& $\ph  8264 \pm   13$ &
& $  5199 ^{+ 124} _{- 124}$
& $  5124 ^{+ 195} _{- 149}$
& $\blank$
& $  5062 ^{+ 173} _{- 173}$ &
& $\cdots$
& $  1462 ^{+  69} _{-  78}$
& $\blank$
\\

KIC010018963
& $\ph  9057 \pm   32$ &
& $  5137 ^{+ 310} _{- 310}$
& $  3820 ^{+ 244} _{- 357}$
& $\blank$
& $\cdots$ &
& $\cdots$
& $  1972 ^{+  55} _{-  56}$
& $\blank$
\\

KIC007976303
& $\ph  9823 \pm   38$ &
& $  7026 ^{+  35} _{-  35}$
& $  6760 ^{+  75} _{-  80}$
& $\blank$
& $\cdots$ &
& $  2835 ^{+  51} _{-  51}$
& $  2671 ^{+  33} _{-  30}$
& $\blank$
\\

KIC011026764
& $\ph  9960 \pm   19$ &
& $  5053 ^{+  33} _{-  33}$
& $  5048 ^{+  43} _{-  38}$
& $\blank$
& $\cdots$ &
& $\cdots$
& $  2316 ^{+  66} _{-  72}$
& $\blank$
\\

KIC011395018
& $ 10548 \pm   22$ &
& $  6504 ^{+ 276} _{- 276}$
& $  7223 ^{+  60} _{-  57}$
& $\blank$
& $\cdots$ &
& $\cdots$
& $  3592 ^{+  40} _{-  38}$
& $\blank$
\\
\enddata
\tablenotetext{1}{In method D actually the acoustic {\em radius}, \tbcz\ was determined, which has been converted to the corresponding acoustic {\em depth}, \taubcz\ using the \tzero\ value derived from the average large separation, \dzz.}
\tablenotetext{2}{A blank cell indicates that an estimation of the acoustic depth was not attempted, and a cell with $\cdots$ indicates that the estimated value did not pass the applied validity check for the relevant method.}
\tablenotetext{3}{Values in parentheses denote a significant detection, but not associated with an acoustic glitch.}
\tablenotetext{4}{For KIC010454113 a secondary value of $\taubcz \approx
1850$\,s is obtained with lesser significance than the value quoted in
methods B and C.}
\tablenotetext{*}{Stars illustrated in Figs.~\ref{fig:res_MJM}--\ref{fig:res_JB}.}

\end{deluxetable*}

We compare the results for all the 19 stars from all four methods in
Table~\ref{tab:res_comp}. The
acoustic radii of the \BCZ\ (\tbcz) determined by method~D have been
converted to acoustic depths (\taubcz) through the relation $\taubcz =
\tzero - \tbcz = (2\dzz)^{-1} - \tbcz$ for comparison with the
values determined by other methods. However, the uncertainties quoted for \taubcz\ from
method D are the intrinsic uncertainties from the method, and do
not include the uncertainties in \tzero. 
Method~D did not consider the presence of the \HIZ\ signal in the data. 

In Fig.~\ref{fig:method_taucomp} we show the 
values of the acoustic depths obtained by different methods for all the
stars. A detailed comparison of the results follows in
Sect.~\ref{sec:discussion_methods_comp}.

\myfigure{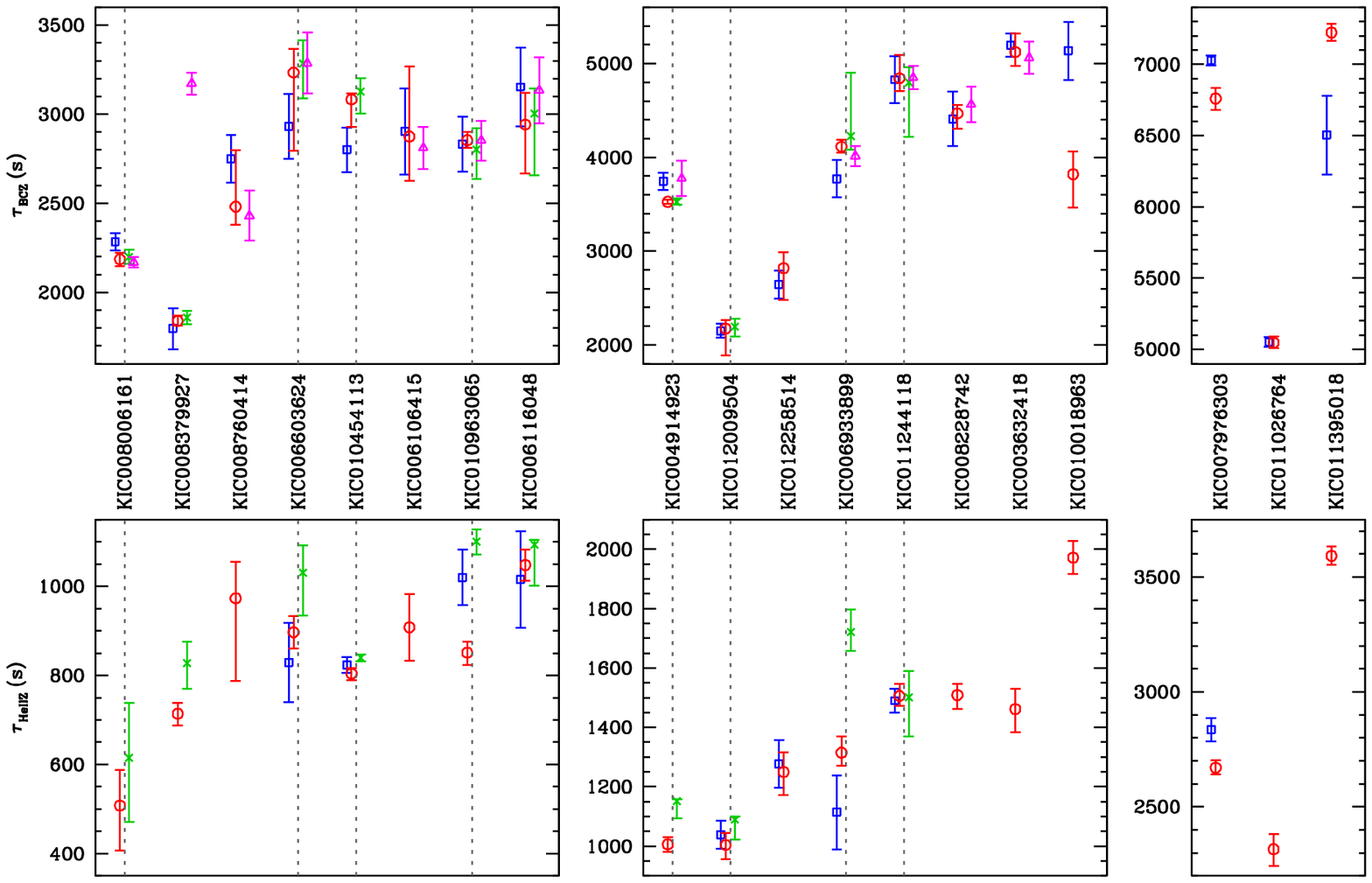}
{
Comparison of acoustic depths \taubcz\ ({\it upper panels}) and \tauhiz\
({\it lower panels}) determined from the
oscillatory signal in the \kep\ frequencies of 19 stars by the four
methods ({\it blue empty squares} for A, {\it red empty
circles} for B, {\it green crosses} for C, and {\it magenta
empty triangles} for D).
The stars have been grouped in three graphs for each glitch with 
different ranges according to their estimated values of \taubcz\ and
\tauhiz.  
The values from different methods for each star are slightly offset
along the horizontal direction from the central positions for clarity. 
The grey dotted lines indicate the selected set of eight stars for which
results are shown in Figs.~\ref{fig:res_MJM} to~\ref{fig:res_JB}.
}
{fig:method_taucomp}
{*}

\section{Comparison with stellar models}
\label{sec:results_models}

The acoustic locations of the glitches determined by all the methods
described above do not depend on detailed modeling of the stars; they
were
derived purely from the observed frequencies. In order to check whether
the estimated values of \taubcz\ and \tauhiz\ are consistent with
typical stellar models, we compared them with values
from representative models of each star. In this, we adopted two
approaches. The first one consisted of comparing the values with a broad
family of models constructed to match the average seismic and
spectroscopic properties of the stars to a fair extent. Additionally, we
compared our values with the theoretical values of one optimally fitted
model for each star.

For this exercise, we compared not the acoustic depths directly, but the
fractional acoustic radii of the \BCZ\ and the \HIZ. This was done for
two reasons. 
Firstly, in a stellar model the acoustic radius of a glitch, 
being calculated from the centre outwards, 
is relatively free from the poorly known 
contribution to the sound speed from the outermost layers of the stars
which lie above the glitch. 
Thus it should be theoretically a more robust quantity than the acoustic
depth, which, being calculated from the surface inwards would include
the sound speed profile in the outer layers. 
However, in converting the acoustic depth estimated from the
oscillatory signal in frequencies to the corresponding acoustic radius
we do use the approximate relationship between the total acoustic radius
and the large separation. Secondly, by considering the fractional
acoustic radii instead of the acoustic radii itself, we remove the
effect of overall homology
scaling of the models at slightly different masses and
radii. Of course, the 
shift between the models
in the relative position of the \BCZ\
and the \HIZ\ inside the star reflects a true departure of the models
from the observed star.
\vskip 40pt

\subsection{Neighborhood models}
\label{sec:results_models_rough}

In this approach, we considered a broad family of theoretical models which
mimic the global properties of the stars and their seismic properties,
as listed in Table~\ref{tab:spec_seismo}, but cannot be claimed to
necessarily have frequencies that match the observed ones very closely.
We deliberately spanned a very broad range in each of the global
properties. This is because our aim here was to only determine the
possible range in the locations of the acoustic glitches in stellar
models similar to the target star, and compare the values estimated
from the \kep\ data. Further, we repeated the exact procedure
of one of the methods (B) to obtain the acoustic depths \taubcz\ and
\tauhiz\ from the theoretical frequencies of the stellar models. This
allowed us also to investigate possible systematic shifts between the
theoretical acoustic locations of the glitches from the models and their
estimated values from the oscillatory signal in the model frequencies
themselves.

We used the Yale Stellar Evolution Code \citep[YREC;][]{Demarque08} to
model the stars. A detailed description of the models can be found in
Appendix~\ref{app:model_details_YREC}.
The first step of our modeling was to use the average large separation
\dzz\ and the frequency of maximum power, \numax,
along with \teff\ and metallicity to determine the masses of the
stars using a grid-based Yale-Birmingham pipeline \citep{Basu10,Gai11}.
\citet{Mathur12} have shown that
grid-based estimates of stellar masses and radii agree very well with
those obtained from more detailed modeling of the oscillation
frequencies of stars, though with slightly lower precision 
\citep[see also][]{Silva11}.

For each star, we specified 8--12 initial masses scanning a
$2\sigma$ range on either side of the mass obtained by the grid
modeling. For each initial mass we modelled the star using three values
of the mixing length parameter $\alpha$ (1.826, 1.7 and 1.5; note
$\alpha=1.826$ is the solar calibrated value of $\alpha$ for YREC). For
each value of mass and $\alpha$ we assumed at least four different values
of the initial helium abundance $Y_0$. In general $Y_0$ ranged from 0.25
to 0.30. The models were evolved from ZAMS and the properties were
output at short intervals to allow us to calculate oscillation
frequencies for the model as it evolved.

All models satisfying the observed constraints on large separation,
small separation, \teff, \logg\ and \febyh\ were selected for
comparison. In order to get a reasonably large number of models the
uncertainty margin was assumed to be 2.0\,\muhz, 2.0\,\muhz, 200\,K, 0.1 dex
and 0.1 dex respectively. The number of models selected with these
criteria ranged from about 80 to 800. For each of these models, the
acoustic depths \taubcz\ and \tauhiz\ were determined by a procedure
exactly similar to method~B adopted for the real data. The theoretically
computed value of the frequency was taken as the mean value and the
uncertainty was adopted as that of the corresponding mode frequency in
the \kep\ data of the concerned star. Thus, the model data set mimicked
the observed data set in terms of number of modes, specific modes used
in the fitting and the error bars on the frequencies.

The comparison of the fractional acoustic radii of the glitches in the
models and our estimates from the \kep\ data are shown in
Fig.~\ref{fig:model_taucomp} for two of the stars in our sample. One of
them (KIC010963065) is a main-sequence star, while the other
(KIC011244118) is in the sub-giant phase. In this figure we show the
fractional acoustic radii of the glitches estimated from the \kep\ data
by method~B along with the values obtained from the frequencies of the
models by the same method, as described above. For the latter, the
typical uncertainty (not shown in the graphs for the sake of clarity) would be
similar to that of the values obtained from the \kep\ data, since we
have assumed the same uncertainties on the theoretical frequencies as
the data. The acoustic depths obtained from the oscillatory signals have
been converted to acoustic radii through the relation $t=\tzero-\tau
\approx (2\dzz)^{-1} - \tau$.  The error bars on the values from these
methods include the uncertainty propagated in the calculation of \tzero\
from \dzz, as listed in Table~\ref{tab:res_comp}. The figure also shows
the theoretical values of the fractional acoustic radii of the glitches
from the models calculated using Eqs.~(\ref{eq:tbcz_thiz})
and~(\ref{eq:tzero}). 
The two illustrated stars are typical of the sample. Results for others
stars are similar to these with the discrepancies between the models and
observations being either smaller or higher in a couple of stars. 
   
\myfigure{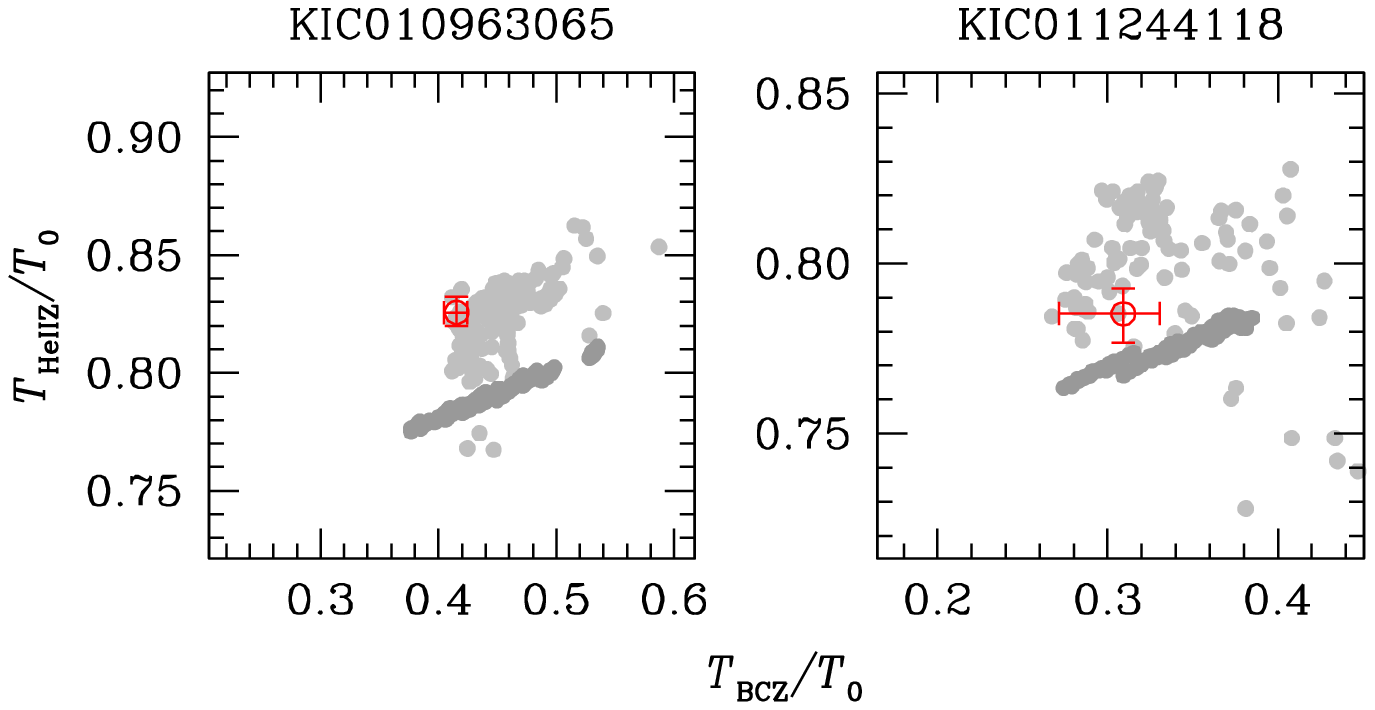}
{
Comparison of fractional acoustic radii $\tbcz/\tzero$ and
$\thiz/\tzero$ determined from the oscillatory signal in the \kep\
frequencies of 2 stars (KIC010963065 and KIC011244118) with corresponding values from representative
models of the stars. In each case the {\it red empty
circle} with error bars represents the results from the data while 
the {\it light grey dots} represent the values determined from the
calculated frequencies of the models. 
Method~B was employed in both cases. 
The {\it dark grey dots} indicate the actual values of the
fractional acoustic radii in the stellar models, calculated using
Eqs.~(\ref{eq:tbcz_thiz}) and~(\ref{eq:tzero}). 
For each star, the ranges of each axis shown in the graph correspond 
to $\pm 1000$\,s and $\pm 500$\,s of the values of \tbcz\ and \thiz\ 
determined from \kep\ data, respectively.
}
{fig:model_taucomp}
{}

\subsection{Optimally fitted models}
\label{sec:results_models_exact}

We also compared our estimated acoustic locations of the glitches to
those in an optimally fitted model of each star. The fitted model was
obtained through the Asteroseismic Modeling Portal
\citep[AMP,][]{Metcalfe09}. The details of the stars studied here can be
found in \citet{Mathur12}. Briefly, the models were obtained by
optimizing the match between the observed frequencies and the modelled
ones. To overcome the issue of the surface effects, we applied the
empirical formula of \citet{Kjeldsen08}.
The model acoustic radii were calculated as per the definitions given in
Eq.~(\ref{eq:tbcz_thiz}).
The comparison of AMP model values of the fractional acoustic radii and
those obtained from the four methods is shown in
Fig.~\ref{fig:amp_taucomp}.

For three of the stars, KIC008006161, KIC006106415 and KIC012009504, 
independent optimized models were made by fitting frequency ratios as 
described by \citet{Silva11,Silva13}.
These models were constructed with the GARSTEC code \citep{Weiss08}, 
and are also shown in Fig.~\ref{fig:amp_taucomp}.
\vskip 20pt

\myfigure{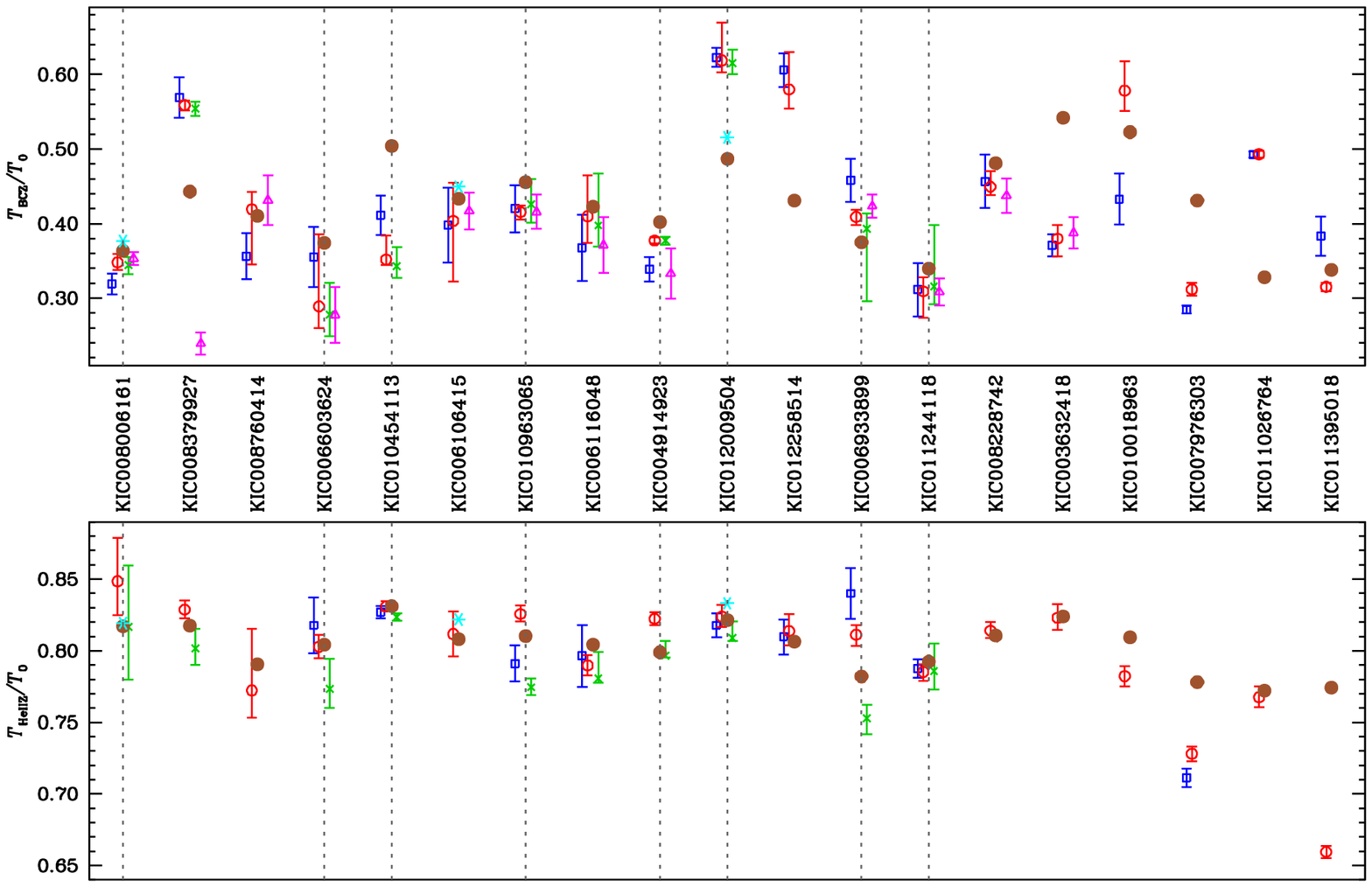}
{Comparison of fractional acoustic radii $\tbcz/\tzero$ and
$\thiz/\tzero$ determined from the
oscillatory signal in the \kep\ frequencies of 19 stars by the four
methods ({\it blue empty squares} for A, {\it red empty
circles} for B, {\it green crosses} for C, and {\it magenta
empty triangles} for D) and AMP model values ({\it brown filled
circles}). For three stars values obtained from 
independently fitted models made by the GARSTEC code are also shown
({\it cyan starred} symbol).
The values from different methods for each star are slightly offset
along the horizontal direction from the central positions for clarity. 
The grey dotted lines indicate the selected set of eight stars for which
results are shown in Figs.~\ref{fig:res_MJM} to~\ref{fig:res_JB}.
}
{fig:amp_taucomp}
{*}

\section{Discussion}
\label{sec:discussion}

\subsection{Comparison between methods and uncertainties}
\label{sec:discussion_methods_comp}

Overall we find remarkably good agreement between the values of the
acoustic depths of the \BCZ\ determined by different methods. In most
cases the \taubcz\ (or the corresponding \tbcz) agree with each other
well within the quoted $1\sigma$ error bars (cf.\
Table~\ref{tab:res_comp} and Fig.~\ref{fig:method_taucomp}). In terms of relative uncertainties, the
\taubcz\ and \tauhiz\ values from different methods agree pairwise within 
10\% and 5\%, respectively for most of the stars. 

The general agreement between \tauhiz\ values from different methods is
not as good as those for \taubcz, although it is mostly within $1\sigma$
uncertainties, and the values match always within 5\% of each other, 
nevertheless.
While comparing our estimates of the acoustic depths \taubcz\ and
\tauhiz\ from methods A, B and C to the acoustic radii from method D, or
from typical models, we invoked the relationship between the total
acoustic radius \tzero\ and the average large separation \dzz, i.e.,
$\tzero \approx (2\dzz)^{-1}$.  This relationship is an approximate one.
Further, the observed average large separation, \dzz, depends on the
adopted frequency range over which it is measured and also on the
adopted method.

Particular attention should be paid to the systematic biases
involved in treating the outermost layers of the star by different
seismic diagnostics dealing with the determination of acoustic depths of
glitches.
Although one may hope that the different methods (seismic diagnostics) will provide similar stellar radii $r$ for the acoustic glitches of both the helium ionization zones and the base of the surface convection zone, the seismically 
measured acoustic depths $\tau(r)$ of these glitches will in general be 
different, for they depend on the very details of the adopted seismic 
diagnostic.
For example, method~C approximates the outer stellar layers by a
polytrope with a polytropic index $m=3.5$, allowing for some account of
the location of the outer turning point of an incident acoustic wave by
approximating the acoustic cutoff frequency $\nu_\mathrm{ac}\simeq
(m+1)/\tau$ \citep{HG07}. The location of the upper turning point
affects the the phase of an incident mode and consequently also the
location of the acoustic glitches, typically increasing their acoustic
depths when the outer layers are approximated by a polytrope
\citep{HG07}.  A (mathematically) convenient way for estimating the
phase at the upper turning point is to relate it to a fiducial location
in the evanescent region far above the upper turning point, where the
mode in the propagating region would ``feel'' the adiabatic sound speed
$c$ to vanish, i.e., where $c=0$.  This location, lying well inside the
evanescent zone of most of the acoustic modes, defines the acoustic
radius or acoustic surface. The squared adiabatic sound speed $c^2$
decreases in the adiabatically stratified region of an outer convection
zone nearly linearly with radius \citep[e.g.][]{Balmforth90,Gough13}. The
location of the acoustic radius can therefore be estimated where the
linearly outward extrapolated $c^2$, with respect to radius $r$,
vanishes.  
In a solar model the so-determined location of the seismic surface is
about 111\,s in acoustic height above the temperature minimum
\citep{Lopes01}, or above 200\,s in acoustic height above the photosphere
\citep{Monteiro94,HG07}. This shift may explain in
part the differences between the location of the acoustic glitches
between the stellar models and our seismically determined values.

There is also a noticeable systematic shift between the acoustic depth 
of the \HIZ\ in a stellar model and its value estimated from the oscillatory 
signal in the model frequencies
(see Fig.~\ref{fig:model_taucomp}).
The fitted value of \tauhiz, as obtained with methods A and B, 
is typically smaller
than the model value by about 100\,s, on
average. 
However, we do not find
such a systematic shift between the \taubcz\ values of the models and
the fits. 
This is a general feature of most of the stars 
and might be because the depth of the HeII ionization zone cannot be
uniquely defined since the ionization zone covers a finite range of
radius. On the other hand, the base of the convection zone is a clearly
defined layer within the star.
Including also the glitch contribution from the first stage of helium
ionization in the seismic diagnostic, additionally to the contribution
from the second stage of helium ionization, perhaps leads to a more accurate
definition of the acoustic depth of helium ionization, as considered in
method C, resulting in larger values for the
fitted acoustic depth \tauhiz\ as indicated in Table~\ref{tab:res_comp}.

In Table~\ref{tab:res_comp} and Figs.~\ref{fig:model_taucomp} and
\ref{fig:amp_taucomp} we did not
take account of these differences in the acoustic depths between the
methods~A, B and C and the calculated stellar models.
It is also worth noticing that the \tzero\ values of AMP models are 
systematically smaller (by about 300\,s, or, by about $4\%$, on average) 
than those estimated from \dzz\ of the \kep\ data
used in this paper (see Table~\ref{tab:results_AMP}). 
This may be due to a combination of factors,
including imperfect modeling of the stars, especially of the surface
layers. 
The AMP fitting has been subjected to the
surface correction technique \citep{Kjeldsen08}, 
while this was not included in the analysis of the present data.
However, even with stellar models, a difference of up to $2\%$
between the \tzero\ values derived from the asymptotic relation
(Eq.~(\ref{eq:tzero})) and that from the large separation is expected
\citep{Hekker13}. 
Nevertheless, this contributes partially to the differences in the
fractional acoustic radii of the glitches 
between AMP models and the estimates from the
data. 
We have not estimated the uncertainties in the model values of the
acoustic radii. Thus it is difficult to make any quantitative comparison
between these values and those obtained from the data.

\begin{deluxetable}{lllll}
\tabletypesize{\small}
\tablecaption{Acoustic radii of glitches in AMP models. Total acoustic radii of the stars in the AMP models and as estimated from the average large separation are also given.
\label{tab:results_AMP}
}
\tablewidth{0pt}

\tablehead{
\colhead{KIC ID}
  & \colhead{\tbcz$^\mathrm{a}$}
  & \colhead{\thiz$^\mathrm{a}$}
  & \colhead{\tzero$^\mathrm{a}$}
  & \colhead{\tzero$^\mathrm{b}$}\\
& \colhead{(s)}
  & \colhead{(s)}
  & \colhead{(s)}
  & \colhead{(s)}
}

\startdata

KIC008006161*
& $  1177$
& $  2647$
& $  3239$
& $  3353 \pm    2$
\\

KIC008379927
& $  1780$
& $  3285$
& $  4019$
& $  4170 \pm    3$
\\

KIC008760414
& $  1675$
& $  3229$
& $  4084$
& $  4273 \pm    3$
\\

KIC006603624*
& $  1629$
& $  3503$
& $  4356$
& $  4549 \pm    4$
\\

KIC010454113*
& $  2326$
& $  3832$
& $  4611$
& $  4757 \pm    9$
\\

KIC006106415
& $  2003$
& $  3735$
& $  4623$
& $  4821 \pm    4$
\\

KIC010963065*
& $  2128$
& $  3787$
& $  4673$
& $  4882 \pm    4$
\\

KIC006116048
& $  2019$
& $  3844$
& $  4779$
& $  4985 \pm    9$
\\

KIC004914923*
& $  2177$
& $  4322$
& $  5410$
& $  5662 \pm    6$
\\

KIC012009504*
& $  2660$
& $  4485$
& $  5461$
& $  5701 \pm    6$
\\

KIC012258514
& $  2761$
& $  5170$
& $  6410$
& $  6711 \pm    9$
\\

KIC006933899*
& $  2488$
& $  5188$
& $  6633$
& $  6963 \pm    9$
\\

KIC011244118*
& $  2286$
& $  5330$
& $  6726$
& $  7012 \pm   19$
\\

KIC008228742
& $  3724$
& $  6278$
& $  7746$
& $  8116 \pm   13$
\\

KIC003632418
& $  4289$
& $  6517$
& $  7911$
& $  8264 \pm   13$
\\

KIC010018963
& $  4516$
& $  6992$
& $  8640$
& $  9057 \pm   32$
\\

KIC007976303
& $  4012$
& $  7240$
& $  9305$
& $  9823 \pm   38$
\\

KIC011026764
& $  3103$
& $  7303$
& $  9459$
& $  9960 \pm   19$
\\

KIC011395018
& $  3363$
& $  7707$
& $  9954$
& $\!\!\! 10548 \pm   22$
\\
\enddata
\tablenotetext{a}{Calculated from AMP model using
Eqs.~(\ref{eq:tbcz_thiz}) and~(\ref{eq:tzero}})
\tablenotetext{b}{Estimated from \dzz\ of \kep\ data}
\tablenotetext{*}{Stars illustrated in Figs.~\ref{fig:res_MJM}--\ref{fig:res_JB}.}

\end{deluxetable}

\subsection{Discussion of specific stars}
\label{sec:discussion_specific}

We discuss further the cases of the eight selected stars in detail below
(ref.~Figs.~\ref{fig:res_MJM}--\ref{fig:res_JB}). The
cases of the 11 remaining stars are somewhat similar to these and a
general understanding of the issues involved may be obtained from the
selected subsample itself.

\subsubsection{KIC008006161}
This is a low mass main sequence star and one of the easiest cases 
for the determination of \taubcz\ or
\tbcz, and all the methods converge to one consistent value which also
agrees with the models. The determination of \tauhiz\ from method~A did
not converge, and hence is is not included. Methods B and C
agree with the model values, as well as the AMP and GARSTEC values.

\subsubsection{KIC006603624}
The oscillatory signal from the \BCZ\ for this star seems to be
substantially weaker
in comparison to the signal from the \HIZ, as borne out in each of
Figs.~\ref{fig:res_MJM}--\ref{fig:res_JB}. In Fig.~\ref{fig:res_MJM},
the amplitude of the oscillations in $\delta\nu_\mathrm{BCZ}$ does not
exceed 0.1\,\muhz, lower than all the other stars shown. It is,
therefore, somewhat difficult to determine the acoustic location of the
\BCZ. This is reflected in the wider and flatter peaks in the histogram
in method B (see Fig.~\ref{fig:res_ABM}), the posterior probability
distribution function (PDF) in method D (see
Fig.~\ref{fig:res_JB}), and the larger error bars from all the four
methods. However, the median values of \taubcz\ derived from all the methods
agree quite well. The values of \tauhiz\ obtained from methods A and B
agree with each other, but only at $2\sigma$ level with that from method
C. The agreement with model values is barely at $1\sigma$ for \tbcz\ but
well within $1\sigma$ for \thiz. The AMP model is also consistent with
the other models.

\subsubsection{KIC010454113}
This star was attempted by the first three methods, and they produce
consistent results for both \BCZ\ and \HIZ. The model values also lie
close to these. This case illustrates the problem of aliasing
encountered while fitting the oscillatory signal due to \BCZ\ (see
Sect.~\ref{app:methods_B}). This is
reflected in the dual peaks in the histogram of method~B
(Fig.~\ref{fig:res_ABM}). Similarly, method~C also finds a secondary
value of \taubcz\ corresponding to the smaller peak from method~B. We
have considered the value with more number of Monte Carlo realisations
as the true value of \taubcz.

\subsubsection{KIC010963065}
All the four methods provide almost identical determinations of \taubcz\
(or corresponding \tbcz), but the \tauhiz\ determined from method~B
agrees with those from the other two methods only at $2\sigma$ level.
However, the model values span the range of the determined \thiz\ (cf.\
Figs.~\ref{fig:model_taucomp} and~\ref{fig:amp_taucomp}).

\subsubsection{KIC004914923}
The \taubcz\ values for this star have been obtained by all the methods, 
and the values agree within 1$\sigma$ uncertainties. The PDF in
method~D shows a very small secondary peak at $\tbcz \sim 3000$\,s, but
the histogram for \taubcz\ in method B shows no such feature. The
\tauhiz\ could not be determined from method~A, but both methods~B and C
provide values which agree within $2\sigma$. 
The derived values for \tbcz\ and
\thiz\ agree quite well with those derived from the model frequencies,
as well as the AMP model values.   

\subsubsection{KIC012009504}
Three methods, A, B and C, were applied for this star.
Although in method~B we observed three peaks in the histogram for \taubcz, they cannot be explained as an aliasing artifact.
Nevertheless, the highest peak value agrees very well with that from methods~A and~C.
The \tauhiz\ values obtained by the three methods also are consistent. 
However, the model values of \tbcz, as derived from both AMP and GARSTEC
models, seem to be smaller than our derived values, and actually correspond to the second highest peak in the histogram from method~B.
Given these facts, it is difficult to ascertain whether the discrepancy
is because of inadequacy of the models to mimic the real star, or due to
failure of our methods to detect the true oscillatory 
signal from the \BCZ.

\subsubsection{KIC006933899}
The \taubcz\ values from all methods agree well within $1\sigma$, as do
the \tauhiz\ determined from methods A and B.
However, \tauhiz\ from method C is more than $2\sigma$
deeper. While the model values of \tbcz\ agree very well with all the
derived values, the \thiz\ from methods A and B match the model fitted
values while that from method C is close to the actual model value. The
AMP model also lies in the vicinity of the other models. 
 
\subsubsection{KIC011244118}
This is a relatively evolved star, for which there are possibly three
mixed dipole modes. Once these modes are removed from consideration, all
the four methods produce exceptionally close median values of both
\taubcz\ (\tbcz) and \tauhiz\ (\thiz), even though the uncertainties are
large in the former.
The model fitted values have a larger scatter possibly because the radial orders of the mixed modes were not identical in such a wide range of models, and thus those were not all removed during fitting.

\subsubsection{Other discrepant cases}
For the remaining 11 stars in the sample, the four methods agree
remarkably well for most cases. Some discrepancies do occur for
KIC008379927, KIC010018963, KIC007976303, and KIC011395018.

For KIC008379927, the first three methods produce consistent results for
\taubcz. Method~D does produce a pronounced peak in the PDF, but at a
value which is unlikely to correspond to the  location of \BCZ. 
A determination of 
\tauhiz\ for this star was only possible from methods B and C which
agree to within $1.5\sigma$.

KIC010018963 seems to be a case where methods A and B have determined the
mutually aliased values of \taubcz, and it is difficult to choose one
over the other as the likely correct location of the \BCZ. 

KIC007976303 is an evolved star for which 23 frequencies could be
used, and even some of these modes could be mixed in nature. Neither the
\taubcz\ nor the \tauhiz\ values determined from methods A and B agree
for this star.

KIC011395018 is a sub-giant star with several mixed modes, and it is
therefore hardly surprising that methods A and B have failed to produce
consistent values of \taubcz. 

\section{Summary}
\label{sec:summary}

We have used the oscillatory signal in the frequencies of 19 \kep\
stars to determine the location of the two major acoustic glitches in
the stellar interior, namely, the base of the convection zone and the
second helium ionization zone. Four independent approaches were used
which exploited either the presence of the signal in the frequencies
themselves, the second differences of the frequencies or the ratio of
small to the large separation. For the stars where more than one method
was applied, we found remarkable agreement in the results, in general.
There were, however, a few discrepant cases, some of which could be traced
to the issue of aliasing between the acoustic depth and radius of the
glitch. In some others, the presence of mixed modes prevented an
accurate analysis. 

As a check on the values of the acoustic radii of the acoustic
glitches determined from the observed frequencies of a star, we also
compared them with theoretical model values. For each star, a large
number of representative models were constructed and the same method
applied for the theoretical frequencies of these as was done for the
observed frequencies. For most of the stars the estimated location of
the glitches were found
to be in close agreement with the model values, within error bars.

Although these results confirm the validity of the analysis done, the major result of the present work is to open the possibility of using the parameters of the glitches to perform model fitting.
The proposed approach adds additional observational constraints, independent from other global and seismic parameters, that can be used in the process of finding the best possible model that reproduces the observations. Most important, these additional constraints can be the source for studying necessary improvements in the physics in order to ensure that all observables are fitted, as has been done for the Sun already \citep[e.g.,][]{JCD11,HG11}.

The present work demonstrates the viability of the techniques applied to
determine the acoustic location of layers of sharp variation of sound
speed in the stellar interior. These methods thus provide powerful tools
for placing constraints on the stratification inside solar-type stars
\citep{Monteiro02, Mazumdar05}. With the availability of precise
frequency sets for a large number of stars from the \kep\ mission, one
can use these techniques to follow the variation of the locations of
acoustic glitches in a large ensemble of solar-type stars populating the
main sequence and sub-giant branches. 

The amplitudes of the oscillatory signals from the acoustic glitches can
also provide useful information about the stellar interior. The location
as well as the strength of the glitch at the base of the convective
envelope in cool stars can place independent constraints on the theories
of convection\citep[e.g.][]{JCD11}. Further, the amplitude of the
oscillatory signal due to the second helium ionization zone can provide
an estimate of the helium content in the stellar envelope
\citep{Perez98,Miglio03,Basu04,mt05,HG07}.

We have demonstrated that the ratio $\taubcz/\tzero$ can be
determined to an accuracy of a few percent. This ratio can be used to
constrain $Z$ and hence may be able to distinguish between different
heavy element mixtures. \citet{vanSaders12} have estimated
that a variation in heavy element abundances by 0.1\,dex can shift the
acoustic depth by about 1\% of the acoustic radius. Thus with improved
data from \kep\ it should be possible to achieve the required accuracy
to distinguish between different heavy element mixtures. There could be
other effects which may also shift the acoustic depth of the convection
zone and these systematic effects need to be studied before we can use
asteroseismic data to study $Z$. 
Nevertheless, the variation in \taubcz\ that we find is consistent with
those of \citet{vanSaders12}.  In view of the large uncertainty in
\febyh\ and
other stellar parameters it is difficult to make a direct
comparison with the predictions of \citet{vanSaders12}.

The analyses carried out here have obviously provided estimates of the
amplitudes and other properties of the fits. The manner in which these
reflect the stellar properties depends sensitively on the details of the
fitting techniques and hence, unlike the acoustic locations, a direct
comparison of those results of the different techniques is not
meaningful. 
The interpretation of
the results in terms of stellar properties will require detailed
comparisons with the results of analysis of stellar model data, to be
carried out individually for each technique. Such studies are envisaged
as a follow-up to the present work.

\acknowledgements
Funding for the \kep\ Discovery mission is provided by 
NASA's Science Mission Directorate. 
AM acknowledges support from the
National Initiative on Undergraduate Science (NIUS) programme of
HBCSE (TIFR). 
SB acknowledges support from NSF grant AST-1105930.
GH acknowledges support by the Austrian Science Fund (FWF) project P21205-N16.
MC and MJPFGM acknowledge financial support from FCT/MCTES, Portugal, through the project {\small PTDC/CTE-AST/098754/2008}.
MC is partially funded by POPH/FSE (EC).
VSA received financial support from the {\it Excellence cluster ``Origin and Structure of the Universe''} (Garching).
DS acknowledges the financial support from CNES.
Funding for the Stellar Astrophysics Centre is provided by The Danish
National Research Foundation. The research is supported by the ASTERISK
project (ASTERoseismic Investigations with SONG and K{\it epler}) funded by the European Research Council (grant agreement no.: 267864).
NCAR is partially supported by the National Science Foundation. This work was partially supported by the NASA grant NNX12AE17G.
This work has been supported, in part, by the European Commission under  SPACEINN (grant agreement {\small FP7-SPACE-2012-312844}).
JB acknowledges Othman Benomar for useful advice on MCMC methods.
We thank Tim Bedding for suggesting improvements to the manuscript.
We thank the anonymous referee for helping us to improve the paper.

\appendix
\section{Methods for fitting the glitches
}
\label{app:methods}

We applied four different methods (labelled A to D) to determine the acoustic locations of the base of the convective zone (BCZ) and the second helium ionization zone (HeIIZ).
Detailed descriptions of these methods are given in the following
sections.
In Figs.~\ref{fig:res_MJM} to \ref{fig:res_JB}, we illustrate each of
the four methods, as applied to some \kep\ stars. 

\subsection{Method A}
\label{app:methods_A}

In this method the strategy is to isolate the signature of the acoustic
glitches in the frequencies themselves.
Compared to the methods involving second differences, described in
Sects.~\ref{app:methods_B} and~\ref{app:methods_C}, this has a different sensitivity to uncertainties
\citep[for a discussion on how the errors propagate, in comparison with 
the uncertainties, when using frequency combinations, please see the works 
by][]{Ballot04,HG07},
does not require having frequencies of consecutive order and avoids the
additional terms that need to be considered when fitting the expression
to frequency differences.  However, it is less robust on the convergence
to a valid solution (the starting guess must be sufficiently close to
the solution), since we do not put any constraints on what we consider to be a smooth component of the frequencies.

Here we assume that only very low-degree data are available so that any
dependence 
of the oscillatory signal
on mode degree can be ignored.  The expression for the
signature due to the \BCZ\ is taken from
\citet{Monteiro94}, \citet{JCD95}, and \citet{Monteiro00}.
The expression for the signature due to the 
\HIZ\ is from \citet{mt98} and \citet{mt05}, with some minor
adaptations and/or simplifications.

In the following $\nu_r$ is a reference frequency, introduced for
normalizing the amplitude.
One good option would be to use $\nu_r=\numax$ (frequency of maximum power), but the actual value selected is not relevant unless we want to compare the amplitude of the signal of different stars.  

The signal from the \BCZ, after removing a smooth component from the
frequencies, is written as (assuming there is no discontinuity in the
temperature gradient at the \BCZ)
\begin{equation}
\delta\nu_\mathrm{BCZ} \simeq A_\mathrm{BCZ} \left( {\nu_r \over \nu}\right)^2
   \sin(4\pi\taubcz\nu + 2\phibcz) \,.
   \label{eq:sig_mjm_bcz}
\end{equation}
The parameters to be determined in this expression are $A_\mathrm{BCZ}$,
\taubcz, \phibcz\ for amplitude, acoustic depth and phase, respectively.
The typical values to expect for a star like the Sun are
$A_\mathrm{BCZ}\sim 0.1$\,\muhz, $\taubcz \sim 2300$\,s and $\phibcz \sim \pi/4$.

The signal from \HIZ, after removing a smooth component from the
frequencies, is described by
\begin{eqnarray}
\delta\nu_\mathrm{\HIZ} & {\simeq} 
& A_\mathrm{\HIZ} \left({\nu_r \over \nu}\right)
   \sin^2 (2\pi\beta_\mathrm{\HIZ}\, \nu) \nonumber\\
& & \times \cos(4\pi\tauhiz\nu + 2\phihiz) \,.
   \label{eq:sig_mjm_hiz}
\end{eqnarray}
The free parameters in this expression are $A_\mathrm{\HIZ}$,
$\beta_\mathrm{\HIZ}$,
\tauhiz, \phihiz\ corresponding, respectively, to amplitude,
acoustic width, acoustic depth and phase.  The typical values to expect
for a solar-like star for these parameters are $A_\mathrm{\HIZ}\sim 1.0$\,\muhz,
$\beta_\mathrm{\HIZ} \sim 130$\,s, $\tauhiz \sim 700$\,s and $\phihiz \sim \pi/4$.
When performing the fitting of Eq.~(\ref{eq:sig_mjm_hiz}), leading to 
these parameters, it must be noted that the signature described by 
Eq.~(\ref{eq:sig_mjm_bcz}) is ignored, being treated as noise in the 
residuals.

If low-degree frequencies are used for a solar-type star, then the signals present in the data correspond to having,
\begin{equation}
\nu = \nu_\mathrm{s1} + \delta\nu_\mathrm{BCZ}
\end{equation}
or
\begin{equation}
\nu = \nu_\mathrm{s2} + \delta\nu_\mathrm{\HIZ}\,.
\end{equation}
Here $\nu_{\mathrm{s}i}$ ($i=1,2$) represents a ``smooth'' component of the mode frequency to be removed in the fitting.
The parameters to fit these expressions are for
Eq.~(\ref{eq:sig_mjm_bcz})
(\BCZ): $A_\mathrm{BCZ}$, \taubcz, \phibcz; and for
Eq.~(\ref{eq:sig_mjm_hiz}) (\HIZ):
$A_\mathrm{\HIZ}$, $\beta_\mathrm{\HIZ}$, \tauhiz, \phihiz.
In this case we treat $\delta\nu_\mathrm{BCZ}$ as noise, since $\nu_\mathrm{s2}$ is treated as including $\nu_\mathrm{s1}$ and $\delta\nu_\mathrm{BCZ}$.

The fitting procedure used is the same method as described by
\citet{Monteiro00}.
This method uses an iterative process in order to remove
the slowly varying trend of the frequencies, leaving the required
signature given in either Eq.~(\ref{eq:sig_mjm_bcz}) or
Eq.~(\ref{eq:sig_mjm_hiz}).  A polynomial in $n$ was fitted to the
frequencies of all modes of given degree $l$ separately, using a
regularized least-squares fit with third-derivative smoothing through a
parameter $\lambda_0$ (see \citet{Monteiro94} for the details).
The residuals to those fits were then fitted for all degrees simultaneously.
The latter fit is the ``signal'' either from the \BCZ\ or from the \HIZ.
This procedure was then iterated (by decreasing the smoothing): at each
iteration we removed from the frequencies the previously fitted signal
and recalculated the smooth component of the frequencies, this time using
the smaller value of $\lambda_j$ ($j=1,2,3,..$).
The iteration converged when the relative changes to the smooth component of the signal fell below $10^{-6}$ (typically this happened for $j\simeq3$).

The initial smoothing, selected through the parameter $\lambda_0$,
defines the range of wavelengths whose variation we want to isolate.
To isolate the signature from the \BCZ, a smaller initial value of
$\lambda_0$ was required in order to ensure that the smooth component
also extracts the signature from the \HIZ.  However, when isolating the
signature from the \HIZ, the shorter wavelength signature from the \BCZ\ was also retained.
We show the fits to Eqs.~(\ref{eq:sig_mjm_bcz}) and
(\ref{eq:sig_mjm_hiz})
in Fig.~\ref{fig:res_MJM} for six of the \kep\ stars that we studied. 

In order to estimate the impact on the fitting of the observational
uncertainties we used Monte Carlo simulations.  We did so by producing
sets of frequencies calculated from the observed values with added
random values calculated from a standard normal distribution multiplied
by the quoted observational uncertainty.
In the fit we removed frequencies with large uncertainties ($> 0.5$\,\muhz\ for the \BCZ\ and $>1.0$\,\muhz\ for the \HIZ).
For each star we produced 500 sets of frequencies, determining the parameters as the standard deviation of the results for the parameters.
Only valid fits were used, as long as the number of valid fits represented more than 90\% of the simulations.
Otherwise we considered that the signature had not been fit successfully
(even if a solution could be found for the observations).

\subsection{Method B}
\label{app:methods_B}

This method involves the second differences of the frequencies to 
determine \taubcz\ and \tauhiz\ simultaneously by fitting a functional
form to the oscillatory signals.

The oscillatory signal in the frequencies due to an acoustic glitch is quite small and is embedded in the frequencies together with a smooth trend 
arising from the regular variation of the sound speed in the stellar interior.
It can be enhanced by using the second differences
\begin{equation}
\Delta_2 \nu(n,l) = \nu(n-1,l) - 2\nu(n,l) + \nu(n+1,l),
\label{eq:2nddiff}
\end{equation}
instead of the frequencies $\nu(n,l)$ themselves
\citep[see, e.g.,][]{Gough90,Basu94,Mazumdar01,Basu04}.

We fitted the second differences to a suitable function representing the oscillatory signals from the two acoustic glitches \citep{Mazumdar01}.
We used the following functional form which has been adapted from \citet{HG07} (Eq.~(22) therein):
\begin{eqnarray}
\label{eq:hou08}
\Delta_2 \nu & = & a_0  \\  
&  + & {b_2\over \nu^2} \; \sin (4\pi\nu\taubcz + 2\phibcz) \nonumber \\
& + & c_0\nu \, e^{-c_2\nu^2} \sin(4\pi\nu\tauhiz + 2\phihiz),\nonumber
\end{eqnarray}
where $a_0$, $b_2$, $c_0$, $c_2$, \taubcz, \phibcz, \tauhiz\ and
\phihiz\ are 8 free parameters of fitting. For one of the stars in our
set (KIC010018963), we needed to use a slightly different function
in which the constant term representing the smooth trend, $a_0$, was 
replaced by a parabolic form: $(a_0 + a_1\nu + a_2\nu^2)$. 
This was necessitated by the sensitivity of the fitted
$\tau$ values to small perturbations to the input frequencies. 
The two $\tau$ values for the \BCZ\ and the \HIZ\ had about 10
times larger (and overlapping) uncertainties with the constant form, as
compared to those with the parabolic form.
Although this
parabolic form could, in principle, be used for all stars as well, it actually
interferes with the slowly varying periodic \HIZ\ component in the limited
range of observed frequencies, and makes it
difficult to determine \tauhiz. Therefore, in adopting
Eq.~(\ref{eq:hou08}) we essentially assumed that the smooth trend in the
frequencies had been reduced to a constant shift in the process of
taking the second differences.
We ignored frequencies which have uncertainties of more than $1\,\muhz$.

While Eq.~(\ref{eq:hou08}) is not the exact form prescribed by
\citet{HG07}, it captures the essential elements of that form while
keeping the number of free parameters relatively small. The ignored
terms can be shown to have relatively smaller contributions.
We note that \citet{Basu04} have shown that the exact form of the
amplitudes of the oscillatory signal does not affect the results
significantly \citep[however, see also][]{HG06}. The fits of
Eq.~(\ref{eq:hou08}) to the second differences
of some selected stars in our sample are shown in the {\em left} panels
of Fig.~\ref{fig:res_ABM}.

The fitting procedure is similar to the one described by
\citet{Mazumdar12}.
The fit was carried out through a nonlinear \chisq\ minimization,
weighted by the uncertainties in the data.
The correlation of uncertainties in the 
second differences was accounted for by defining the \chisq\ using
a covariance matrix.  The effects of the uncertainties were considered by
repeating the fit for 
$1000$ realizations of the data, produced by perturbing the 
frequencies by random uncertainties corresponding to a normal
distribution with standard deviation equal to the quoted $1\sigma$
uncertainty in the frequencies.
The successful convergence of such a non-linear fitting
procedure is somewhat dependent on the choice of reasonable initial
guesses. To remove the effect of initial guesses affecting the final
fitted parameters, we carried out the fit for multiple combinations of
starting values.
For each realization, the fitting was repeated for $100$ random combinations of initial guesses of the free parameters
and the fit which produced the minimum value of \chisq\ was accepted.

The median value of each parameter for 1000 realizations was taken as its
fitted value.  
The $\pm 1\sigma$ uncertainty in the parameter was estimated
from the range of values covering $34\%$ area about the median in the
histogram of fitted values, assuming the error distribution to be Gaussian.
Thus the quoted uncertainties in these parameters reflect the width of these histograms on two sides of the median value.
The histograms for \taubcz\ and \tauhiz\ and the ranges of
initial guesses of the corresponding parameters are shown in the
{\em right} panels of Fig.~\ref{fig:res_ABM}.

In some cases the fitting of the \taubcz\ parameter suffers from the
aliasing problem \citep{Mazumdar01}, where a significant fraction of the
realizations are fitted with \taubcz\ equal to
$\tilde{\tau}_\mathrm{BCZ} \equiv \tzero - \taubcz$. 
This becomes apparent from the
histogram of \taubcz\ which appears bimodal with a reflection around
$\tzero/2$ (e.g., for KIC010454113, shown in Fig.~\ref{fig:res_ABM}). 
In such cases, we chose the higher peak in the histogram to
represent the true value of \taubcz, and the uncertainty in the parameter was calculated after ``folding'' the histogram about the acoustic mid-point
$\tzero/2$, the acoustic radius being estimated from the mean large
separation, \dzz. For a few stars there are multiple peaks in the
histogram for \taubcz, not all of which can be associated with the true
depth of the \BCZ\ or its aliased value (e.g., for KIC012009504, shown in 
{\em bottom right} panel of Fig.~\ref{fig:res_ABM}). 
In such cases we chose only the most prominent peak to determine the median and the uncertainty was estimated from the width of that peak.
The remaining realizations were also neglected for estimating the other parameters such as \tauhiz.

\subsection{Method C}
\label{app:methods_C}

Approximate expressions for the frequency contributions $\delta\nu_i$ arising from acoustic glitches in solar-type stars were recently 
presented by \citet{HG07, HG11}, which we adopt here for producing
the results presented in Fig.~\ref{fig:res_GH} and
Table~\ref{tab:res_comp}.
A detailed discussion of the method can be found in those references; we
therefore present here only a summary.
The complete expression for $\delta\nu_i$ is given by 
\begin{equation}
\delta\nu_i=\delta_\gamma\nu_i+\delta_{\rm c}\nu_i+\delta_{\rm u}\nu_i\, ,
\label{eq:delnu}
\end{equation}
where the terms on the RHS are the individual acoustic glitch components 
located at different depths inside the star.

The first component,
\begin{eqnarray}
\delta_\gamma\nu&=&-\sqrt{2\pi}A_{\rm II}\Delta^{-1}_{\rm II}
\left[\nu+\textstyle\frac{1}{2}(m+1)\Delta_0\right]\cr
&
\times&\hspace{-8pt} \Bigl[\tilde\beta\int_0^{\tzero}\kappa^{-1}
{\rm e}^{-{(\tau-\tilde\eta\tau_{\rm II})^2 \over 2\tilde\mu^2\Delta^2_{\rm II}}}|x|^{1/2}
|{\rm Ai}(-x)|^2\,{\rm d}\tau\cr
&+&\hspace{-8pt}\int_0^{\tzero}\hspace{-8pt}\kappa^{-1}
{\rm e}^{-{(\tau-\tau_{\rm II})^2\over 2\Delta^2_{\rm II}}}|x|^{1/2}
|{\rm Ai}(-x)|^2\,{\rm d}\tau\Bigr]
\label{eq:delgamnu}
\end{eqnarray}
arises from the variation in $\gamma_1$ induced by 
helium ionization.
The constant $m=3.5$ is a representative polytropic index in the
expression for the approximate effective phase $\psi$ appearing in the
argument $x={\rm sgn}(\psi)|3\psi/2|^{2/3}$ of the Airy function ${\rm Ai}(-x)$. 
We approximate $\psi$ as
\begin{equation}
\psi(\tau)=\kappa\omega\tilde\tau-(m+1)\cos^{-1}[(m+1)/\omega\tilde\tau]\,,
\end{equation}
if $\tilde\tau>\tau_{\rm t}$, and
\begin{equation}
\psi(\tau)=|\kappa|\omega\tilde\tau-(m+1)\ln[(m+1)/\omega\tilde\tau+|\kappa|]\,,
\end{equation}
if $\tilde\tau{\le}\tau_{\rm t}$, in which 
$\tilde\tau{=}\tau{+}\omega^{-1}\epsilon_{\rm II}$, with $\epsilon_{\rm II}$ 
($\epsilon_{\rm I}{=}\epsilon_{\rm II}$) being a phase
constant, and $\tau_{\rm t}$ is the location of the upper turning point of 
the mode.
The location of the upper turning point of the oscillation mode is
determined approximately from the polytropic representation of the acoustic
cutoff frequency, leading to the expressions 
$\kappa(\tau)=[1-(m+1)^2/\omega^2\tilde\tau^2]^{1/2}\,$.

The coefficients associated with the glitch contribution from the first stage of 
helium ionization (\ion{He}{1})
(first integral expression in Eq.~(\ref{eq:delgamnu})) 
are related to the coefficients of the second stage of helium ionization (\ion{He}{2}) 
by the constant ratios $\tilde\beta:=A_{\rm I}\Delta_{\rm II}/A_{\rm II}\Delta_{\rm I}$,
$\tilde\eta:=\tau_{\rm I}/\tau_{\rm II}$ and $\tilde\mu:=\Delta_{\rm I}/\Delta_{\rm II}$,
where $A_{\rm II}$ is the amplitude factor of the oscillatory \HeII\ glitch component, 
$\Delta_{\rm II}$ is the acoustic width of the glitch and $\tau_{\rm II} \equiv \tauhiz$ 
is its acoustic depth beneath the seismic surface. For the constant ratios 
$\tilde\beta$, $\tilde\eta$, and $\tilde\mu$ we set the values 0.45, 0.70 and 0.90,
respectively, as did \citet{HG07, HG11}.
With this approach the addition of the \ion{He}{1} contribution does not introduce additional fitting coefficients in the seismic diagnostic~(\ref{eq:delgamnu}).

The second component in Eq.~(\ref{eq:delnu}),
\begin{eqnarray}
\delta_{\rm c}\nu&\simeq&A_{\rm c}\Delta_0^3\nu^{-2}
   \left(1+1/16\pi^2\tau_0^2\nu^2\right)^{-1/2}\cr
&&\hspace{-3pt}\times\Big\{\cos[2\psi_{\rm
c}+\tan^{-1}(4\pi\tau_0\nu)]\nonumber \\
&&\hspace{12pt}-(16\pi^2\tilde{\tau}_\mathrm{BCZ}^2\nu^2\!+\!1)^{1/2}
\Big\}\, 
\label{eq:delcnu}
\end{eqnarray}
arises from the acoustic glitch at the base of the convection zone resulting from a near discontinuity (a true discontinuity in theoretical models using local mixing length theory with a non-zero mixing length at the lower boundary of the convection zone) in the second derivative of density.
We model this acoustic glitch with a discontinuity in the squared
acoustic cutoff frequency $\omega_{\rm c}^2$ at \taubcz\,, with $A_{\rm c}$ 
being proportional to the jump in $\omega_{\rm c}^2$, coupled with an exponential
relaxation to a putative, glitch-free, model in the radiative zone beneath, 
with a relaxation time scale $\tau_0=80$\,s, as did \citet{HG07, HG11}. 
This leads to
\begin{eqnarray} 
\psi_{\rm c}&=&\kappa_{\rm c}\omega\tilde{\tau}_\mathrm{BCZ}\cr
&-& (m+1)\cos^{-1}\left[(m+1)/\tilde{\tau}_\mathrm{BCZ}\omega\right]\cr
&+&\pi/4\,,
\end{eqnarray}
where $\kappa_{\rm c}=\kappa(\taubcz)$ and 
$\tilde{\tau}_\mathrm{BCZ}=\taubcz+\omega^{-1}\epsilon_{\rm c}$
with $\epsilon_{\rm c}$ being a constant phase.

The additional upper-glitch component $\delta_{\rm u}\nu_i$ ($i$
enumerates individual frequencies), which is
produced, 
in part, by wave refraction in the stellar core, by the ionization
of hydrogen and by the upper superadiabatic
boundary layer of the envelope convection zone, is difficult to model.
We approximate its contribution to $\Delta_2\nu$ as a series of inverse 
powers of $\nu$, truncated at the cubic order:
\begin{equation}
\Delta_2\delta_{\rm u}\nu_i=\sum_{k=0}^3a_k\nu_i^{-k}\, .
\label{eq:delsmooth}
\end{equation}

The eleven coefficients 
$\eta_\alpha=(A_{\rm II}$, $\Delta_{\rm II}$, $\tauhiz$, 
$\epsilon_{\rm II}$, $A_{\rm c}$, $\taubcz$, $\epsilon_{\rm c},
a_0, a_1, a_2, a_3),\, \alpha=1,...,11,$ 
were found by fitting the second differences (cf.\
Eq.~(\ref{eq:2nddiff})),
\begin{eqnarray}
\Delta_{2i}\nu(n,l)&{:=}&\nu(n-1,l)-2\nu(n,l)+\nu(n+1,l)\nonumber\\[5pt]
&&\hspace{-48pt}
\simeq\;\Delta_{2i}(\delta_\gamma\nu+\delta_{\rm c}\nu+\delta_{\rm u}\nu)
=: g_i(\nu_j;\eta_\alpha)
\label{eq:secdiff}
\end{eqnarray}
to the corresponding observations by minimizing
\begin{equation}
E_{\rm g}=\sum_{i,j}(\Delta_{2i}\nu-g_i)C^{-1}_{\Delta ij}(\Delta_{2j}\nu-g_j)\,,
\label{eq:minsecdiff}
\end{equation}
where $C^{-1}_{\Delta ij}$ is the $(i,j)$ element of the inverse of the 
covariance matrix {C$_\Delta$} of the observational uncertainties in 
$\Delta_{2i}\nu$, computed, perforce, under the assumption that the 
uncertainties in the frequency data $\nu_i$ are independent. 
The covariance matrix $C_{\eta\alpha\gamma}$ of the uncertainties in the
fitting coefficients $\eta_\alpha$ were established by Monte Carlo simulation.

Results for six of the stars in our sample are shown
in Fig.~\ref{fig:res_GH} in which also the individual acoustic glitch
contributions are illustrated in the {\it lower} panels.

\subsection{Method D}
\label{app:methods_D}

Determinations of acoustic depths of glitches are biased by surface
effects \citep[e.g.][]{JCD95}. A way to remove such biases is to
consider acoustic radii of the signatures. First, it is possible to
perform posterior determinations of radii by comparing depths to the
total acoustic radius of the star derived from the average large
separation
\dzz\ \citep[see][]{Ballot04}; it is also possible to directly
measure acoustic radii by considering the small separations $d_{01}$ and
$d_{10}$ or the frequency ratios $r_{01}$ and $r_{10}$ as shown by
\citet{RV03}.

We use the 3-point differences:
\begin{equation}\label{eqn:d01}
d_{01,n}=\frac{1}{2}(2\nu_{n,0}-\nu_{n-1,1}-\nu_{n,1}),
\end{equation}
\begin{equation}\label{eqn:d10}
d_{10,n}=-\frac{1}{2}(2\nu_{n,1}-\nu_{n,0}-\nu_{n+1,0})
\end{equation}
and the corresponding ratios:
\begin{eqnarray}\label{eqn:rat}
r_{01,n}=\frac{d_{01,n}}{\Delta\nu_{1,n}},& r_{10,n} = \frac{d_{10,n}}{\Delta\nu_{0,n+1}}.
\end{eqnarray}
We denote by $d_{010}$ and $r_{010}$ the sets $\{d_{01},d_{10}\}$ and $\{r_{01},r_{10}\}$, respectively.

Using these variables, the main contributions of outer layers are
removed \citep[see][]{Rox05}. The global trend of these variables 
then gives information on the core of the star
\citep[e.g.][]{Silva11,Cunha11}. Nevertheless, the most internal
glitches, such as the \BCZ, also imprint their signatures over the global
trend.  Using solar data, \citet{Rox09} showed that we can recover the
acoustic radius of the \BCZ\ ($T_{\mathrm{BCZ}}$) by the use of a Fourier
transform on the residuals obtained after removing the global trend.  As
a consequence of this approach, information about surface layers,
including \ion{He}{1} and \ion{He}{2} ionization zones, are lost.

We used an approach similar to \citet{Rox09} and develop a semi-automatic
pipeline which extracts glitches from a frequency table of $l=0$ and
1 modes. Instead of fitting a background first, then making a Fourier
transform, we do both simultaneously by fitting the variable
$y=\nu^*r_{010}$ (or $y=\nu^*d_{010}$) with the following expression:
\begin{equation}
f(\nu)= \sum_{k=0}^m \frac{c_k}{(\nu+\nu_\mathrm{r})^k} + A\sin(4\pi T \nu + \phi),\label{eq:modd01}
\end{equation}
where $\nu^* = \nu / \nu_\mathrm{r}$ and $\nu_\mathrm{r}$ is a reference frequency. We
have $m+3$ free parameters: $\{c_k\}$, $A$, $T$, and $\phi$.

To be able to estimate reliable uncertainties for \tbcz\ we performed a
Markov Chain Monte Carlo (MCMC) to fit the data. Our MCMC fitting
algorithm is close to the one described, for example, in \citet{Benomar09}
or \citet{HandbergCampante11}, but without parallel tempering. Moreover,
in our case the noise is not multiplicative and exponential, but
additive and normal.  We fixed $m=2$ and $\nu_\mathrm{r} = 0.8 \numax$,
similar to the value adopted for the Sun by \citet{Rox09}.  We
used uniform priors for $A$, $T$ and $\phi$. The prior for $A$ was very
broad (between 0 and $10\max(|y|)$); $\phi$ was $2\pi$-periodic with a
uniform prior over $[0,2\pi]$, and $T$ was restricted to $[\delta T,
\tzero-\delta T]$, where $\delta T = (\max(\nu)-\min(\nu))^{-1}$.
To determine priors for $\{c_k\}$, we first performed a standard linear
least-square fitting without including the oscillatory component. Once
we obtained the fitted values $\tilde c_k$ and associated uncertainties $\sigma_k$,
we used for $c_k$ priors which are uniform over $[\tilde c_k-
3\sigma_k,\tilde c_k+ 3\sigma_k]$ with Gaussian decay around this
interval.  After a  200\,000-iteration learning phase \citep[the
learning method we used is based on][]{Benomar09}, MCMC was run for $10^7$
iterations. We tested the convergence by verifying the consistence of
results obtained by using the first and second half of the chain.

Posterior probability distribution functions (PDF) for $T$ are plotted for
different stars in Fig.~\ref{fig:res_JB}.  Some PDFs exhibit a unique and
clear mode that we attributed to the \BCZ. \tbcz\ was then estimated
as the median of the distribution and 1$\sigma$ error bars were
taken as 68\,\% confidence limit of the PDF.  Nevertheless, some
other PDFs exhibit other small peaks. They can be due to noise, but they
also can be due to other glitch signatures present in the data,
especially tiny residuals from the outer layers. To take care of such
situations, we also computed \tbcz\ by isolating the highest
peak and fitting it with a Gaussian profile. When the secondary peaks
are small enough, the values and associated uncertainties obtained through the two methods are very similar.
We have been successful mainly on main-sequence stars,
while subgiant stars presenting mixed modes strongly disturb
the analysis.

\section{Details of YREC stellar models}
\label{app:model_details_YREC}

The input physics for the YREC models included the OPAL equation of state
\protect{\citep{rn02}} and the OPAL high-temperature opacities \citep{IR96}
supplemented with low-temperature opacities from \citet{Ferguson05}.
All nuclear reaction rates were from \citet{Adelberger98}, except for the rate of the $^{14}\mathrm{N}(p,\gamma)^{15}\mathrm{O}$ reaction, which was fixed at the value of \citet{Formicola04}.
Core overshoot of $0.2H_p$ was included where relevant.  
For the stars that did not have very low metallicities,
we included the diffusion and settling of helium and heavy elements.
This was done as per the prescription of \citet{Thoul94}.

We use four grids for this: the models from the Yonsei-Yale (YY)
isochrones \citep{Demarque04}, and the grids of
\citet{Dotter08}, \citet{Marigo08} and \citet{Gai11}.



\begin{thebibliography}{}
{\small 

\bibitem[Adelberger et al.(1998)]{Adelberger98}
Adelberger E.~G., Austin, S.~M., Bahcall, J.~N., et al.\ 
1998, Rev. Mod. Phys., 70, 1265

\bibitem[Appourchaux et al.(2012)]{Appourchaux12} 
Appourchaux, T., Chaplin, W.~J., Garc{\'{\i}}a, R.~A., et al.\ 
2012, \aap, 543, A54

\bibitem[Asplund et al.(2009)]{Asplund09}
Asplund, M., Grevesse, N., Sauval, A.~J., \& Scott, P.\
2009, \araa, 47, 481

\bibitem[Ballot et al.(2004)]{Ballot04} 
Ballot, J., Turck-Chi{\`e}ze, S., \& Garc{\'{\i}}a, R.~A.\ 
2004, \aap, 423, 1051

\bibitem[Balmforth \& Gough(1990)]{Balmforth90}
Balmforth N.~J.~\& Gough D.~O.\ 
1990, \apj, 362, 256

\bibitem[Basu(1997)]{Basu97} 
Basu, S.\ 
1997, \mnras, 288, 572 

\bibitem[Basu \& Antia(2008)]{Basu08} 
Basu, S., \& Antia, H.~M.\ 
2008, \physrep, 457, 217

\bibitem[Basu \ea(1994)Basu, Antia, \& Narasimha]{Basu94}
Basu, S., Antia, H.~M., \& Narasimha, D.\
1994, \mnras, 267, 209

\bibitem[Basu et al.(2010)]{Basu10}
Basu, S., Chaplin, W.~J., Elsworth, Y.\ 
2010, \apj, 710, 1596

\bibitem[Basu et al.(2004)]{Basu04} 
Basu, S., Mazumdar, A., Antia, H.~M., \& Demarque, P.\ 
2004, \mnras, 350, 277 

\bibitem[Bedding et al.(2010)]{Bedding10} 
Bedding, T.~R., Kjeldsen, H., Campante, T.~L., et al.\ 
2010, \apj, 713, 935

\bibitem[Benomar et al.(2009)]{Benomar09} 
Benomar, O., Appourchaux, T., \& Baudin, F.\ 
2009, \aap, 506, 15 

\bibitem[Borucki et al.(2010)]{Borucki10}
Borucki W.~J.\ et al.\ 
2010, Sci, 327, 977

\bibitem[Bruntt et al.(2012)]{Bruntt12} 
Bruntt, H., Basu, S., Smalley, B., et al.\ 
2012, \mnras, 423, 122 

\bibitem[Christensen-Dalsgaard et al.(1995)Christensen-Dalsgaard, Monteiro, \& Thompson]{JCD95}
Christensen-Dalsgaard, J., Monteiro, M.~J.~P.~F.~G., Thompson, M.~J.\  
1995, \mnras, 276, 283

\bibitem[Christensen-Dalsgaard et al.(2011)Christensen-Dalsgaard, Monteiro, Rempel, \& Thompson]{JCD11}
Christensen-Dalsgaard, J., Monteiro, M.~J.~P.~F.~G., Rempel, M., Thompson, M.~J.\  
2011, \mnras, 414, 1158

\bibitem[Cunha \& Brand{\~a}o(2011)]{Cunha11} 
Cunha, M.~S., \& Brand{\~a}o, I.~M.\ 
2011, \aap, 529, A10 

\bibitem[Demarque et al.(2004)]{Demarque04}
Demarque, P., Woo, J.-H., Kim, Y.-C., \& Yi, S.~K.\ 
2004, ApJS, 155, 667

\bibitem[Demarque et al.(2008)]{Demarque08} 
Demarque, P., Guenther, D.~B., Li, L.~H., Mazumdar, A., \& Straka, C.~W.\ 
2008, \apss, 316, 31 

\bibitem[Dotter et al.(2008)]{Dotter08}
Dotter, A., Chaboyer, B., Jevremovic, D. et al.\ 
2008, \apj, 178, 89

\bibitem[Ferguson et al.(2005)]{Ferguson05}
Ferguson, J.~W., Alexander, D.~R., Allard, F., et al.\ 
2005, \apj, 623, 585

\bibitem[Formicola et al.(2004)]{Formicola04}
Formicola A., Imbriani, G., Costantini, H. et al.\ 
2004, Physics Letters B, 591, 61

\bibitem[Gai et al.(2011)]{Gai11}
Gai, N., Basu, S., Chaplin, W.~J., Elsworth, Y.\ 
2011, \apj, 730, 63

\bibitem[Garc{\'{\i}}a et al.(2011)]{Garcia11}
Garc{\'{\i}}a, R.~A., Hekker, S., Stello, D., et al.\ 
2011, \mnras, 414, L6

\bibitem[Gilliland et al.(2010)]{Gilliland10}
Gilliland, R.~L., Jenkins, J.~M., Borucki, W.~J., et al.\ 
2010, \apj, 713, L160

\bibitem[Gough(1986)]{Gough86}
Gough D.~O.\ 
1986, in Hydrodynamic and Magnetodynamic Problems in the Sun and Stars.
Y.~Osaki, ed., Univ.\ Tokyo Press, 117

\bibitem[Gough(1990)]{Gough90} 
Gough, D.~O.\ 
1990, Progress of Seismology of the Sun and Stars, 367, 283 

\bibitem[Gough(2002)]{Gough02}
Gough D.~O.\ 
2002, in Proceedings of the First Eddington Workshop on Stellar
Structure and Habitable Planet Finding. B.\ Battrick, F.\ Favata, I.~W.\
Roxburgh \& D.\ Galadi, eds., ESA SP-485, Noordwijk: ESA, p.\ 65

\bibitem[Gough(2013)]{Gough13} 
Gough, D.\ 2013, 
\solphys, 287, 9

\bibitem[Gough \& Sekii(1993)]{Gough93}
Gough, D.~O., \& Sekii, T.\ 
1993, in GONG 1992: Seismic Investigation of the Sun and Stars. T.~M.\
Brown, ed., ASP Conf.\ Ser.\ Vol.\ 42., Astron.\ Soc.\ Pac., San Francisco, p.\ 177

\bibitem[Gough \& Thompson(1988)]{Gough88}
Gough, D.~O., \& Thompson, M.~J.\ 
1988, in Advances in Helio- and Asteroseismology. J.\ Christensen-Dalsgaard, 
S.\ Frandsen, eds., Reidel Publishing Company, 155

\bibitem[Grevesse \& Sauval(1998)]{GS98}
Grevesse, N., \& Sauval, A.~J.\
1998, \ssr, 85, 161

\bibitem[Handberg \& Campante(2011)]{HandbergCampante11}
Handberg, R., \& Campante, T.~L.\ 
2011, \aap, 527, A56

\bibitem[Hekker et al.(2013)]{Hekker13}
Hekker, S., Elsworth, Y., Basu, S., et al.\ 
2013, accepted for publication in \mnras\ (arXiv:1306.4323)

\bibitem[Houdek(2004)]{Houdek04}
Houdek, G.\ 
2004, in Equation-of-state and Phase-transition Issues in Models of
Ordinary Astrophysical Matter. V.\ Celebonovic, W.\ D{\"a}ppen, D.~O.\
Gough, eds., AIP Conf.\ Proc., Vol.\ 731, AIP, New York, p.\ 193

\bibitem[Houdek \& Gough(2006)]{HG06}
Houdek, G., Gough, D.~O.\ 
2006, in Proc.\ SOHO18/GONG 2006/HELAS~I: Beyond the spherical Sun.
M.~Thompson, K.~Fletcher, eds., ESA SP-624, Noordwijk, p.\ 88

\bibitem[Houdek \& Gough(2007)]{HG07}
Houdek, G., Gough, D.~O.\ 
2007, \mnras, 375, 861

\bibitem[Houdek \& Gough(2011)]{HG11}
Houdek, G., Gough, D.~O.\ 
2011, \mnras, 418, 1217

\bibitem[Iglesias \& Rogers(1996)]{IR96}
Iglesias, C.~A., Rogers, F.~J.\ 
1996, \apj, 464, 943

\bibitem[Kjeldsen et al.(2008)]{Kjeldsen08} 
Kjeldsen, H., Bedding, T.~R., \& Christensen-Dalsgaard, J.\ 
2008, \apjl, 683, L175

\bibitem[Koch et al.(2010)]{Koch10}
Koch D.~G.\ et al.\ 
2010, \apj, 713, L79

\bibitem[Lopes \& Gough(2001)]{Lopes01}
Lopes I.~P.~\& Gough D.~O.\ 
2001, \mnras, 322, 473  

\bibitem[Marigo et al.(2008)]{Marigo08}
Marigo, P., Girardi, L., Bressan, A. et al.\ 
2008, \aap, 482, 883

\bibitem[Mathur et al.(2012)]{Mathur12}
Mathur, S., Metcalfe, T.~S., Woitaszek, M. et al.\ 
2012, \apj, 749, 152

\bibitem[Mazumdar(2005)]{Mazumdar05} 
Mazumdar, A.\ 
2005, \aap, 441, 1079 

\bibitem[Mazumdar \& Antia(2001)]{Mazumdar01} 
Mazumdar, A., \& Antia, H.~M.\ 
2001, \aap, 377, 192 

\bibitem[Mazumdar et al.(2011)]{Mazumdar11} 
Mazumdar, A., Michel, E., Antia, H.~M., Deheuvels, S.\  
2011, in 
Transiting planets, vibrating stars and their connection,
Proceedings of the Second {\it CoRoT} Symposium. A.~Baglin,
M.~Deleuil, E.~Michel, C.~Moutou \& T.~Seman, eds.,
p.\ 197

\bibitem[Mazumdar et al.(2012)]{Mazumdar12} 
Mazumdar, A., Michel, E., Antia, H.~M., Deheuvels, S.\  
2012, \aap, 540, A31

\bibitem[Metcalfe et al.(2009)]{Metcalfe09} 
Metcalfe, T.~S., Creevey, O.~L., \& Christensen-Dalsgaard, J.\ 
2009, \apj, 699, 373 

\bibitem[Miglio et al.(2003)]{Miglio03}
Miglio A., Christensen-Dalsgaard J., di Mauro M.~P., Monteiro 
M.~J.~P.~F.~G., Thompson M.~J.\ 
2003, in Asteroseismology Across the HR Diagram, Proceedings of the 
Asteroseismology Workshop, Porto, Portugal. M.~J.~Thompson, M.~S.~Cunha,
M.~J.~P.~F.~G. Monteiro, eds., Kluwer Academic Publishers, Dordrecht,
p.\ 537

\bibitem[Miglio et al.(2010)]{Miglio10} 
Miglio, A., Montalb{\'a}n, J., Carrier, F., et al.\ 
2010, \aap, 520, L6 

\bibitem[Monteiro \ea(1994)Monteiro, Christensen-Dalsgaard, \& Thompson]{Monteiro94}
Monteiro, M.~J.~P.~F.~G., Christensen-Dalsgaard, J., Thompson, M.~J.\
1994, \aap, 283, 247

\bibitem[Monteiro et al.(2000)Monteiro, Christensen-Dalsgaard, \& Thompson]{Monteiro00} 
Monteiro, M.~J.~P.~F.~G., Christensen-Dalsgaard, J., Thompson, M.~J.\ 
2000, \mnras, 316, 165

\bibitem[Monteiro et al.(2002)Monteiro, Christensen-Dalsgaard, \& Thompson]{Monteiro02} 
Monteiro, M.~J.~P.~F.~G., Christensen-Dalsgaard, J., Thompson, M.~J.\ 
2002, in Stellar Structure and Habitable Planet Finding. F.~Favata,
I.~W.~Roxburgh, D.~Galad\`{i}-Enr\'{i}quez, eds., ESA Publications
Division, ESA SP-485, 291

\bibitem[Monteiro \& Thompson(1998)]{mt98}
Monteiro, M.~J.~P.~F.~G., \& Thompson, M.~J.\ 
1998, in Structure and Dynamics of the Interior of the Sun and Sun-like Stars. 
S.~G.~Korzennik, A.~Wilson, eds., ESA Publications Division, ESA SP-418, 819

\bibitem[Monteiro \& Thompson(2005)]{mt05}
Monteiro, M.~J.~P.~F.~G., \& Thompson, M.~J.\ 
2005, \mnras, 361, 1187

\bibitem[Morel \& Lebreton(2008)]{ml08}
Morel, P., \& Lebreton, Y.\
2008, \apss, 316, 61


\bibitem[P\'erez Hern\'andez \& Christensen-Dalsgaard(1998)]{Perez98} 
P\'erez Hern\'andez, F., \& Christensen-Dalsgaard, J.\ 
1998, \mnras, 295, 344

\bibitem[Piau et al.(2005)]{Piau05} 
Piau, L., Ballot, J., \& Turck-Chi{\`e}ze, S.\ 
2005, \aap, 430, 571 

\bibitem[Rogers \& Nayfonov(2002)]{rn02}
Rogers, F.~J., Nayfonov, A.\ 
2002, ApJ, 576, 1064

\bibitem[Roxburgh(2005)]{Rox05} 
Roxburgh, I.~W.\ 
2005, \aap, 434, 665

\bibitem[Roxburgh(2009)]{Rox09} 
Roxburgh, I.~W.\ 
2009, \aap, 493, 185 

\bibitem[Roxburgh(2011)]{Rox11} 
Roxburgh, I.~W.\ 
2011, in Transiting planets, vibrating stars and their connection,
Proceedings of the Second {\it CoRoT} Symposium, A.~Baglin,
M.~Deleuil, E.~Michel, C.~Moutou \& T.~Seman, eds.,
p.\ 161

\bibitem[Roxburgh \& Vorontsov(1994)]{RV94}
Roxburgh, I.~W.~\& Vorontsov, S.~V.\
1994, \mnras, 268, 880

\bibitem[Roxburgh \& Vorontsov(2003)]{RV03}
Roxburgh, I.~W.~\& Vorontsov, S.~V.\
2003, \aap, 411, 215

\bibitem[Silva Aguirre et al.(2011)]{Silva11} 
Silva Aguirre, V., Ballot, J., Serenelli, A.~M., \& Weiss, A.\ 
2011, \aap, 529, A63 

\bibitem[Silva Aguirre et al.(2013)]{Silva13}
Silva Aguirre, V., Basu, S., Brand{\~a}o, I.~M., et al.\ 
2013, \apj, 769, 141

\bibitem[Tassoul(1980)]{Tassoul80}
Tassoul M.\ 
1980, \apjs, 43, 469

\bibitem[Thoul et al.(1994)]{Thoul94}
Thoul A.~A., Bahcall J.~N., Loeb A.\ 
1994, ApJ, 421, 828

\bibitem[Weiss \& Schlattl(2008)]{Weiss08} 
Weiss, A., \& Schlattl, H.\ 
2008, \apss, 316, 99 

\bibitem[van Saders \& Pinsonneault(2012)]{vanSaders12} 
van Saders, J.~L., \& Pinsonneault, M.~H.\ 
2012, \apj, 746, 16

\bibitem[Verner et al.(2011)]{Verner11} 
Verner, G.~A., Elsworth, Y., Chaplin, W.~J., et al.\ 
2011, \mnras, 415, 3539 

\bibitem[Vorontsov(1988)]{Vorontsov88}
Vorontsov, S.~V.\ 
1988, in Advances in Helio- and Asteroseismology. J.~Christensen-Dalsgaard,
S.~Frandsen, eds., (Reidel), IAU Symp, 123, 151

}
\end{thebibliography}
\end{document}